\documentclass[letterpaper,11pt]{article}
\usepackage{jheppub} 
\pdfoutput=1
\usepackage{amsmath}
\usepackage{xcolor}
\usepackage{youngtab}
\usepackage{subcaption}
\usepackage{graphicx}
\usepackage{verbatim}
\usepackage{framed}
\usepackage{slashed} 
\definecolor{shadecolor}{rgb}{0.90,0.90,0.90}
\def\beq{\begin{eqnarray}}\def\eeq{\end{eqnarray}}
\def\be{\begin{equation}}\def\ee{\end{equation}}
\def\g{\gamma}

\def\m{\mu}
\def\n{\nu}
\def\a{\alpha}

\def\b{\beta}
\def\d{\delta}

\def\D{\Delta}

\def\tf{\tilde{f}}

\def\3s{{s \choose 3}}
\def\4s{{s \choose 4}}
\def\5s{{s \choose 5}}
\def\6s{{s \choose 6}}

\def\12{\dfrac{1}{2}}

\def\2{\ell_2}

\def\be{\begin{equation}}
\def\ee{\end{equation}}
\def\bea{\begin{eqnarray}}
\def\eea{\end{eqnarray}}
\def\ba{\begin{array}}
\def\ea{\end{array}}
\def\bec{\begin{center}}
\def\ec{\end{center}}

\def\g{\gamma}

\def\m{\mu}
\def\n{\nu}
\def\a{\alpha}

\def\b{\beta}
\def\d{\delta}

\def\D{\Delta}

\def\tf{\tilde{f}}

\def\3s{{s \choose 3}}
\def\4s{{s \choose 4}}
\def\5s{{s \choose 5}}
\def\6s{{s \choose 6}}

\def\12{\dfrac{1}{2}}

\def\2{\ell_2}

\def\be{\begin{equation}}
\def\ee{\end{equation}}
\def\bea{\begin{eqnarray}}
\def\eea{\end{eqnarray}}
\def\ba{\begin{array}}
\def\ea{\end{array}}
\def\bec{\begin{center}}
\def\ec{\end{center}}

\def\a{\alpha} 
\def\b{\beta}  
\def\g{\gamma} 

\def\d{\delta} 
\def\D{\Delta}

\def\m{\mu}
\def\n{\nu}

\def\tf{\tilde f}

\newcommand{\Z}{\mathbb{Z}}

\def\tilde{\widetilde}
\def\hat{\widehat}
\def\bar{\overline}

\def\ts{{\mathtt s}}

\def\tf{{\mathtt f}}

\newcommand{\tyng}{\tiny\yng}

\definecolor{nicered}{rgb}{0.7,0.1,0.1}
\definecolor{nicegreen}{rgb}{0.1,0.5,0.1}

\def\denom{{\mathtt D}}
\definecolor{mGreen}{rgb}{0,0.6,0}
\definecolor{mgray}{rgb}{0.6,0.6,0.6}
\definecolor{mpurple}{rgb}{0.58,0,0.82}
\definecolor{backgroundColour}{rgb}{0.95,0.95,0.92}

\definecolor{mred}{rgb}{0.5,0.0,0.0}
\definecolor{mgreen}{rgb}{0.0,0.4,0.0}
\definecolor{mblue}{rgb}{0.0,0.0,0.6}
\definecolor{myellow}{rgb}{0.4,0.4,0.0}
\definecolor{mpink}{rgb}{0.4,0.0,0.4}
\definecolor{mcyan}{rgb}{0.0,0.4,0.4}
\definecolor{mblack}{rgb}{0.0,0.0,0.0}

\def\g{{\gamma}}

\def\a{{\alpha}}
\def\b{{\beta}}

\def\d{{\delta}}

\newcommand{\bd}[1]{\begin{fmffile}{#1}\begin{fmfgraph*}}
		\newcommand{\ed}{\end{fmfgraph*}\end{fmffile}}

\def\0{{(0)}}
\def\1{{(1)}}
\def\2{{(2)}}
\def\3{{(3)}}
\def\4{{(4)}}

\def\+{{(+)}}
\def\-{{(-)}}

\def\be{\begin{equation}}
\def\ee{\end{equation}}
\def\beq{\be\begin{array}{c}}
	\def\eeq{\end{array}\ee}

\numberwithin{equation}{section}

\newcommand{\mf}[1]{\mathfrak #1}

\abstract{
We count the number of independent solutions to crossing constraints of four point functions involving charged scalars and charged fermions in a CFT with large gap in the spectrum. To find the CFT data we employ recently developed analytical functionals to charged fields. We compute the corresponding higher dimensional flat space S matrices in an independent group theoretic manner and obtain agreement with our CFT counting of ambiguities. We also write down the local lagrangians explicitly. Our work lends further evidence to \cite{Heemskerk:2009pn} that any CFT with a large charge expansion and a gap in the spectrum has an AdS bulk dual.}

\title{\bf Bulk locality for scalars and fermions with global symmetry}
\date{}

\author{\!\!\!\! Subham Dutta Chowdhury$^{a,}$\footnote{subham@theory.tifr.res.in}, Kausik Ghosh$^{b,}$\footnote{kau.rock91@gmail.com}\\ ~~~~\\
\it $^{a}$ Department of Theoretical Physics,\\ \it Tata Institute of Fundamental Research,\\ \it Mumbai 400005. \\
\it ${^b}$Centre for High Energy Physics,
\it Indian Institute of Science,\\ \it C.V. Raman Avenue, Bangalore 560012, India. }
\begin{document}
\preprint{TIFR/TH/21-7}

\maketitle

\section{Introduction}

AdS/CFT duality \cite{Maldacena:1997re} is a paradigm shift in our understanding of quantum gravity. This duality relates a conformal field theory to a theory of gravity in one higher dimension. In a strict sense there are no local observables in theory of quantum gravity and all the dynamics are encoded in the boundary (hologram). Though in an effective description of classical gravity, the notion of locality is well defined as is obvious in our universe.  Therefore emergence of bulk locality is a natural question that people have been curious about from the beginning days of AdS/CFT. Here we revisit the ideas introduced in \cite{Heemskerk:2009pn} and look for further evidence in support of this claim. There are two natural length scales in the problem, the AdS radius $R$ and the string length  $l_s$. Now if we consider an effective field theory in AdS the locality is supposed to hold down till $l_s$. Since $R=\lambda^{1/4} l_s$ and $\lambda=g^2N$ where $N$ is the number of color in the boundary conformal gauge theory and $g$ is the coupling constant, it is expected that, for local description $\lambda$ has to be very large so that $R$ is parametrically larger than $l_s$. AdS/CFT \cite{Maldacena:1997re, Gubser:1998bc, Witten:1998qj}, therefore implies that the dimension of the operators, dual to string excitations, will have large dimensions \footnote{AdS/CFT dictionary tells us $\Delta(\Delta-d)=m^2R^2$}. So a local bulk theory in AdS will correspond to a perturbative CFT in large central charge $(c\sim N^2)$ expansion with large gap in the spectrum, i.e., all single trace operators with spin greater than $2$ will have large dimensions.

In \cite{Heemskerk:2009pn}, the authors made a remarkable conjecture that these implications run in the opposite direction as well, a CFT with a large gap in the spectrum necessarily is described by a local bulk dual \footnote{Throughout the text we will denote the bulk dimensions as $D$, while the boundary dimensions will be labelled by $d$}. They studied the crossing equation of four point correlator of a single trace operator, ${\cal O}$, in a theory containing just the single trace operator itself and its double traces in large central charge expansion (to order $O\left( c^{-1} \right)$). There are as many independent solutions to the crossing equations as there are local bulk counterterms with the specified support in spin. The independent data of the crossing is encoded in the anomalous dimensions of the double trace operators ${\cal O}_{n,l} = {\cal O} \square^{n}\partial_{\mu_1}\partial_{\mu_2}\cdots \partial_{\mu_l}  {\cal O}- \text{traces}$. The number of undetermined anomalous dimensions due to crossing are in agreement with the number of local counterterms for a given support in spin. The consistency of the crossing equation therefore requires the existence of a local bulk dual. The explicit expressions for the anomalous dimensions have also been computed for explicit bulk counter terms \cite{Heemskerk:2009pn, Heemskerk:2010ty}. In \cite{Caron-Huot:2021enk} the authors have proved the assumptions of \cite{Heemskerk:2009pn} from CFT axioms for the case of identical scalars (see also \cite{Kundu:2021qpi} which arrived at similar bounds motivated from Regge boundedness of the corresponding Lorentzian CFT correlator). In this present paper we provide more evidence for this conjecture by considering ($d=4$) scalars and ($d=1$) fermions charged under global symmetry. The results for scalars can be extended to any dimensions using the analytical functions discussed in subsection \ref{cbauf}. 

We consider  holographic CFTs of coloured scalars (in $d=4$ initially) charged under the fundamental and adjoint representations of $SO(N)$ and $SU(N)$ respectively. Following \cite{Heemskerk:2009pn, Heemskerk:2010ty}, we solve the crossing equation of four point correlators to order $O(c^{-1})$  or $O\left( N^{-2} \right)$ with finite support over spin $L$. We note that the parameter $N$ in the large $N$ expansion of the correlator is different from the $N$ of $SO(N)$ and $SU(N)$ respectively. We find that for a fixed spin $L$ support, we can encode the number of undetermined parameters for $O\left(c^{-1}\right)$ crossing in terms of partition functions. A comprehensive list of such partition functions are given in table \ref{spinsuppscintro}.

\begin{table}[h!]
	\begin{center}
		\begin{tabular}{|c|c|c|} 
			\hline
			\textbf{Group} & \textbf{Representations} & \textbf{Spin support: Even and Odd spin}\\
			\hline
			$SO(N)$& Fundamental & $\frac{1}{8} (L+2) (3 L+4),~~~\frac{1}{8} (L+1) (3 L+5)$ \\
			\hline
			$SO(N)$& Adjoint & $\frac{1}{4} (L+2) (3 L+4),~~~\frac{1}{4} (L+1) (3 L+5)$ \\
			\hline
			$SU(N)$& Fundamental & $\frac{3}{4} L(L+2)+1,~~~\frac{3}{4} (L+1)^2$ \\
			\hline
			$SU(N)$& Adjoint & $\frac{1}{4} (L+2) (3 L+4),~~~\frac{1}{4} (L+1) (3 L+5)$ \\
			\hline
		\end{tabular}
		\caption{Spin support for bulk contact terms: Scalars with global symmetry}\label{spinsuppscintro}
	\end{center}
\end{table}

We also verify this counting by bootstrapping large $c$ CFTs using analytic functionals \cite{Mazac:2019shk, Caron-Huot:2020adz}. Construction of analytic functionals in CFTs have been of interest in recent years \cite{El-Showk:2016mxr, Mazac:2018mdx, Mazac:2018ycv,Paulos:2019gtx, Paulos:2019fkw, Paulos:2020zxx}. In \cite{Mazac:2019shk}, the authors construct analytic functionals for CFTs in $d>1$ and relate them to Regge bounded Witten diagrams and in \cite{Caron-Huot:2020adz} it was shown that for holographic CFTs consisting of just scalars and its double traces, such functionals reproduce the counting of bulk counterterms in terms of anomalous dimensions. A remarkable achievement is that, the finite spin support is not an assumption in their analysis but rather an outcome in trying to bootstrap in this method. We extend their analysis to construct functionals for scalar correlators with global symmetry and verify the counting of undetermined anomalous dimensions obtained using usual bootstrap methods.

We also consider flat space S-matrices in $D=2$ for massive Majorana fermions charged under global symmetry which are dual to $d=1$ fermions in CFT. In order to solve the crossing equation, we write down the functionals and the necessary subtractions for $d=1$ fermions charged under global symmetry following \cite{Paulos:2020zxx, Ghosh:2021ruh}. The ambiguities obtained can, similar to the scalars, be encoded in terms of a partition function however with the vital difference from the higher dimensional case in the sense that the support is over derivatives rather than spin  (See table \ref{spinsuppfintro} where exponent of $x$ denotes order of derivatives derivatives).                

\begin{table}[h!]
	\begin{center}
		\begin{tabular}{|c|c|c|} 
			\hline
			\textbf{Group} & \textbf{Representations} & \textbf{Derivative support}\\
			\hline
			$-$& $-$ & $x^{4b+2}$ \\
			\hline
			$SO(N)$& Fundamental & $(1+2x^2)x^{4b}$ \\
			\hline
		\end{tabular}
		\caption{Bulk contact terms ($D=2$): Fermions with and without global symmetry}\label{spinsuppfintro}
	\end{center}
\end{table}

In order to evaluate the bulk contact interaction, we follow the techniques of \cite{Chowdhury:2019kaq, Chowdhury:2020ddc} to group theoretically evaluate the ``local module"
for the four point scalar and fermions charged under global symmetry. As defined in \cite{Chowdhury:2019kaq}, local modules are in one-to-one correspondence with local bulk Lagrangians and are graded by order of derivatives and $S_3$ transformation properties ($S_2$ for $D=2$ fermion S-matrices).

\begin{table}[h!]
	\begin{center}

		\begin{tabular}{|c|c|c|} 
			\hline
			\textbf{S-matrix Lagrangian} & \textbf{$S_3$ representations} & \textbf{Module structure}\\
			\hline
			$L_{SO(N),f}$& ${\bf 3}$ & $\sum_{m,n}  a_{m,n} \prod_{b=1}^m\prod_{c=1}^n\left(\partial_{\m_b} \partial_{\n_c} \phi_i \phi_i \right)  \left(\partial^{\m_b} \phi_j \partial^{\n_c} \phi_j \right)$ \\
			\hline
			$L^1_{SO(N),a}$& ${\bf 3}$ & $ \sum_{m,n}  a_{m,n} \prod_{b=1}^m\prod_{c=1}^n \left(\partial_{\m_b} \partial_{\n_c} \phi_{ij} \phi_{ji} \right)  \left(\partial^{\m_b} \phi_{kl} \partial^{\n_c} \phi_{lk} \right)$ \\
			\hline
			$L^2_{SO(N),a}$& ${\bf 3}$ & $\sum_{m,n}  a_{m,n} \prod_{b=1}^m\prod_{c=1}^n \left(\partial_{\m_b} \partial_{\n_c} \phi_{ij} \partial^{\m_b} \phi_{jk}  \phi_{kl} \partial^{\n_c} \phi_{li} \right)$ \\
			\hline
			$L^1_{SU(N),a}$& ${\bf 3}$ & $\sum_{m,n}  a_{m,n} \prod_{b=1}^m\prod_{c=1}^n \left(\partial_{\m_b} \partial_{\n_c} \phi^i_j \phi^j_i \right)  \left(\partial^{\m_b} \phi^k_l \partial^{\n_c} \phi^l_k \right)$ \\
			\hline
			$L^2_{SU(N),a}$& ${\bf 3}$ & $\sum_{m,n}  a_{m,n} \prod_{b=1}^m\prod_{c=1}^n\left(\partial_{\m_b} \partial_{\n_c} \phi^i_j \partial^{\m_b} \phi^j_k  \phi^k_l \partial^{\n_c} \phi^l_i \right)$ \\
			\hline
			$L_{SU(N),f}$& ${\bf 6}$ & $\sum_{m,n}  a_{m,n} \prod_{b=1}^m\prod_{c=1}^n \left(\partial_{\m_b} \partial_{\n_c} \phi_i \partial^{\n_b}\bar{\phi}^i \right)  \left( \phi_j  \partial^{\m_b}\bar{\phi}^j \right)$ \\
			\hline
		\end{tabular}
		\caption{Generators of scalar local module with global symmetry}\label{lagrangianintro}
	\end{center}
\end{table}

 We obtain the following module of bulk scalar Lagrangians (Table \ref{lagrangianintro}). In this table the entry in the LHS denotes the details of the local Lagrangian (the subscript denotes the lie group and the irreducible representation: $f$ for fundamental and $a$ for adjoint), the middle column denotes the $S_3$ transformation property and the right most column denotes the explicit generator upto a given order ($2m+2n$) in derivatives.
 For (bulk) dimensions $D>3$, we also evaluate the support of the scalar flat space S-matrices (and equivalently the bulk Lagrangians) over spin. This can be encoded in the form of a partition function for every bulk Lagrangian corresponding to an irreducible representation of $S_3$. This can be summarised by the following table \ref{spinsupportintro}. In this table the entries in the first two columns have the same significance as table \ref{lagrangianintro} and the right most column denotes the number of independent Lagrangians contributing to a spin support $L$.     

\begin{table}[h!]
	\begin{center}
		\begin{tabular}{|c|c|c|} 
			\hline
			\textbf{S-matrix Lagrangian} & \textbf{$S_3$ representations} & \textbf{Spin support}\\
			\hline
			$L_{SO(N),f}$& ${\bf 3}$ & $n_{I_1}(L)+n_{I_2}(L)+n_{I_3}(L)$ \\
			\hline
			$L^1_{SO(N),a}$& ${\bf 3}$ & $n_{I_1}(L)+n_{I_2}(L)+n_{I_3}(L)$ \\
			\hline
			$L^2_{SO(N),a}$& ${\bf 3}$ & $n_{I_1}(L)+n_{I_2}(L)+n_{I_3}(L)$ \\
			\hline
			$L^1_{SU(N),a}$& ${\bf 3}$ & $n_{I_1}(L)+n_{I_2}(L)+n_{I_3}(L)$ \\
			\hline
			$L^2_{SU(N),a}$& ${\bf 3}$ & $n_{I_1}(L)+n_{I_2}(L)+n_{I_3}(L)$ \\
			\hline
			$L_{SU(N),f}$& ${\bf 6}$ & $n_{I_1}(L)+2n_{I_2}(L)+2n_{I_3}(L)+ n_{I_4}(L)$ \\
			\hline
		\end{tabular}
		\caption{Spin support for scalar S-matrices}\label{spinsupportintro}
	\end{center}
\end{table}
where \begin{eqnarray}
n_{I_1}(L)&=&\frac{1}{2} \left(\left\lfloor \frac{L}{2}\right\rfloor +1\right) \left(\left\lfloor \frac{L}{2}\right\rfloor +2\right),\qquad
n_{I_2} (L) = \frac{1}{2} \left(\left\lfloor \frac{L-1}{2}\right\rfloor +1\right) \left(\left\lfloor \frac{L-1}{2}\right\rfloor +2\right) \nonumber\\
n_{I_3} (L) &=& \frac{1}{2} \left(\left\lfloor \frac{L-2}{2}\right\rfloor +1\right) \left(\left\lfloor \frac{L-2}{2}\right\rfloor +2\right),\qquad
n_{I_4} (L) = \frac{1}{2} \left(\left\lfloor \frac{L-3}{2}\right\rfloor +1\right) \left(\left\lfloor \frac{L-3}{2}\right\rfloor +2\right)\nonumber\\
\end{eqnarray} 

For the case of coloured fermions in $D=2$, we explicitly evaluate the module of local Lagrangians charged under the fundamental of $SO(N)$ and obtain a perfect match with the number of functional ambiguities at a particular derivative order. The Lagrangians are listed in table \ref{tablefermions}.

\begin{table}[h!]
	\begin{center}
		\begin{tabular}{|c|c|} 
			\hline
			\textbf{S-matrix Lagrangian}  & \textbf{Module structure}\\
			\hline
			$L^F$&  $	\sum_{m,n}  a_{m,n} \prod_{b=1}^m\prod_{c=1}^n(\partial_{\m_b} \partial_{\n_c}\bar{\psi}\partial^{\m_b}\psi)( \partial^{\n_c}\bar{\psi}\psi)$ \\
			\hline
			$L^F_{SO(N),f}$& $\sum_{m,n}  a_{m,n} \prod_{b=1}^m\prod_{c=1}^n(\partial_{\m_b} \partial_{\n_c}\bar{\psi}_i \partial^{\m_b}\psi_j)(\partial^{\n_c}\bar{\psi}_j\psi_i)$\\
			\hline
		\end{tabular}
		\caption{Generators of fermion local module w/o global symmetry}\label{tablefermions}
	\end{center}
\end{table}


The paper is organised as follows, we obtain the solutions to the crossing equation at $O\left(c^{-1}\right)$ with finite support in spin for identical scalars ($d>2$) with colour ($SO(N)$ and $SU(N)$ fundamental and adjoint) in section \ref{swgsdg4}. We find independent group theoretic agreement of scalar bulk contact terms with finite support in spin in subsection \ref{cfssm} and also list out the explicit flat-space Lagrangians for the same. We analyse the large $N$ CFT crossing equations in terms of analytic functionals in subsection \ref{cbauf} and find agreement with results obtained using usual bootstrap. In section \ref{mfi1dwgs}, we study the solutions of crossing of $d=1$ fermions charged under fundamental of $SO(N)$ using analytic functionals and derive the number of contact terms required as a function order by order in derivatives. We match the counting in subsection \ref{mffssi11d} by an independent method of evaluating and explicitly writing down Lagrangians which generate flat-space Majorana fermion S-matrices in $D=2$ bulk.   
 





\section{Scalars with global symmetry in $d= 4$}\label{swgsdg4}

In this section we consider crossing of four point functions of scalars and its double traces charged under a global symmetry group at large central charge \cite{Heemskerk:2009pn}. For concreteness we consider scalars charged under the fundamental and adjoint of $SO(N)$ and $SU(N)$ respectively but in principle these methods can be applied to any lie group. The theory we consider has no other single trace operator and hence the spectrum consists of just the single trace coloured scalar and its double traces. We write down the crossing equation (with finite support in spin) along with the different types of double trace operators being exchanged. We consider the bulk local operators (charged under the same global symmetry group) and evaluate the most general contact terms with derivatives.

\subsection{Constraints from crossing $SO(N)$ }
Let us consider the four point function of  identical scalars transforming under fundamental representation of $SO(N)$,
\begin{equation}
\langle\phi_i(x_1)\phi_j(x_2)\phi_k(x_3)\phi_\ell(x_4)\rangle=\frac{\mathcal{G}_{ijk\ell}(u,v)}{x_{12}^{2\Delta_{\phi}}x_{34}^{2\Delta_{\phi}}}=\sum_R \left(t^{(R)}\right)^{ijk\ell}\frac{G^{(r)}(u,v)}{x_{12}^{2\Delta_{\phi}}x_{34}^{2\Delta_{\phi}}},
\end{equation}
where the cross ratios are given by,
\begin{equation}
u=\frac{x_{12}^2x_{34}^2}{x_{13}^2x_{24}^2},\,\,\,\,\, v=\frac{x_{14}^2x_{23}^2}{x_{13}^2x_{24}^2},
\end{equation}
and  $\left(t^{(R)}\right)^{ijk\ell}$ are given by,
\begin{equation}\label{sonce}
\begin{split}
\left(t^{(S)}\right)^{ijk\ell} =&\delta_{ij}\delta_{k\ell}, \qquad 
\left(t^{(T)}\right)^{ijk\ell} =\left(\frac{\delta_{ik}\delta_{j\ell}+\delta_{i\ell}\delta_{jk}}{2}-\frac{1}{N}\delta_{ij}\delta_{k\ell}\right),\qquad
\left(t^{(A)}\right)^{ijk\ell} =\frac{\delta_{ik}\delta_{j\ell}-\delta_{i\ell}\delta_{jk}}{2}.\\
\end{split}
\end{equation}
The indices $R$ run over the labels which represent the irreducible structures which appear in the tensor product of two fundamentals of $SO(N)$, we choose to label them by (Singlet ($S$), Traceless symmetric ($T$), Anti-symmetric ($A$)) corresponding to the following irreducible representations, in terms of $SO(N)$ young tableaux.
$$\left({\mathbb I}, {\tyng(2)}, {\tyng(1,1)}\right)$$

 The crossing symmetry implies the following constraints \cite{Li:2015rfa, Rattazzi:2010yc, Vichi:2011ux, Poland:2011ey, Kos:2015mba, Kos:2013tga},
\begin{eqnarray}\label{crossinggsonf}
\mathcal{G}_{ijk\ell}(u,v)=\left(\frac{u}{v}\right)^{\Delta_{\phi}}\mathcal{G}_{kji\ell}(v,u), \qquad
G^{(\mu)} (u,v)= M_{SO(N),f}^{\mu \nu} \left(\frac{u}{v}\right)^{\Delta_{\phi}} G^{(\nu)} (v,u),
\end{eqnarray}
where $\mu=1,2,3$ denotes $S,T$ and $A$ respectively and the crossing matrix $M_{SO(N),f}$ is given in \eqref{cmsonf}. Now we have conformal block decomposition for each sector of the correlator,
\begin{equation}
G^{(R)}(u,v)=\sum_{\Delta,\ell}C^R_{\Delta,\ell} ~ g_{\Delta,\ell}(u,v),
\end{equation}
where R could stand for singlet(S), traceless symmetric(T) or antisymmetric(A) sectors. Note that throught this section we assume that $N$ is large enough so that we can ignore specific representations which might appear for low enough $N$. We show in later in this section, that this counting works for $SO(4)$ where we have additional  structures in the crossing. The double trace operators corresponding to the different irreducible sectors are as follows,
\begin{eqnarray}\label{dtsonf}
{\cal O}^{(S)}_{n,l}= \phi_i\square^{n}\partial^l\phi_i, \qquad {\cal O}^{(T)}_{n,l}= \phi_{(i}\square^{n}\partial^l\phi_{j)}- \frac{\delta_{ij}}{N}\phi_{k}\square^{2n}\partial^l\phi_{k}, \qquad {\cal O}^{(A)}_{n,l}= \phi_{[i}\square^{n}\partial^l\phi_{j]}. \nonumber\\
\end{eqnarray}
Note that the $S$ and $T$ sectors can take on only even values of spin and the $A$ sector has support only over odd spins. We now study the crossing equation in a large central charge ($c$) expansion.
\begin{eqnarray}\label{largeNsc}
A^R(z,{\bar z}) &=& A^R_0 (z,{\bar z}) + \frac{1}{c} A^R_1 (z,{\bar z}) + \cdots ,\qquad C^R(n,\ell_R)= C^R_0(n,\ell_R) + \frac{1}{c} C^R_1 (n,\ell_R) + \cdots,\nonumber\\
 \Delta^R(n,\ell_R)&=&\Delta^R_0(n,\ell_R) + \frac{1}{c} \gamma^R_1 (n,\ell_R) + \cdots  \nonumber\\
\end{eqnarray} 

We want to solve \eqref{crossinggsonf}, order by order in $c$ assuming the ansatz \eqref{largeNsc}. Schematically,
\begin{eqnarray}\label{crossing2sc}
A^R_0 (z,{\bar z}) &=& 1  + \sum_{n=0}^\infty \sum_{\ell=0}^\infty C^R_0(n,\ell_R) g_{2\Delta_\phi+2n+\ell_R,\ell_R} (z, \bar{z}), \nonumber\\
A^R_1 (z,{\bar z})&=& \sum_{n=0}^\infty \sum_{\ell=0}^\infty \left(C^R_1(n,\ell_R)  + C^R_0(n,\ell_R) \gamma_1(n,\ell_R) \frac{\partial}{\partial n}  \right) g_{2\Delta_\phi+2n+\ell_R ,\ell_R} (z, \bar{z}).
\end{eqnarray}   


\subsubsection*{Solution for $O\left(c^0\right)$}
The Mean field amplitude is given by,
\begin{equation}
\langle \phi_i(x_1)\phi_j(x_2)\phi_k(x_3)\phi_{\ell}(x_4) \rangle=\delta_{ij} \delta_{kl} 1 +\delta_{i k}\delta_{j \ell} u^{\Delta_{\phi}}+\delta_{il} \delta_{jk} \left(\frac{u}{v}\right)^{\Delta_{\phi}}.
\end{equation}

Decomposing it into irreducible sectors one finds,
\begin{equation}\label{N0sonf}
\begin{split}
& G^{(S)}(u,v)=1+\frac{1}{N}\Bigg(u^{\Delta_{\phi}}+\left(\frac{u}{v}\right)^{\Delta_{\phi}}\Bigg),\qquad G^{(T)}(u,v)=\Bigg(u^{\Delta_{\phi}}+\left(\frac{u}{v}\right)^{\Delta_{\phi}}\Bigg),\\
& G^{(A)}(u,v)=\Bigg(u^{\Delta_{\phi}}-\left(\frac{u}{v}\right)^{\Delta_{\phi}}\Bigg).
\end{split}
\end{equation}
Assuming that the conformal block decompositions are given by ,
\begin{equation}\label{confdesonf}
G^{(R)}(u,v)=\sum_{\Delta,\ell}C^{(R)}_{\Delta,\ell} g_{\Delta,\ell}(u,v),
\end{equation}
we can solve for $C^{(R)}_{\Delta,\ell}$. In order to do so, we use \eqref{crossing2sc} and the explicit form of the $d=4$ blocks listed below. 

\begin{eqnarray}\label{d4confblocks}
g_{\Delta,\ell} = \frac{z {\bar z}}{z- {\bar z}}\left( \kappa\left(\Delta+\ell,z\right)\kappa\left(\Delta-\ell-2,{\bar z}\right) -\kappa\left(\Delta+\ell,{\bar z}\right)\kappa\left(\Delta-\ell-2,z\right) \right), \qquad
\kappa\left(x,y\right)=y^{\frac{x}{2}}\, _2F_1(x,x;2 x;y).\nonumber\\
\end{eqnarray}

We expand \eqref{N0sonf} and \eqref{confdesonf}  in the limit $z \rightarrow 0, \bar z \rightarrow 0$ and solve order by order to get\footnote{Note that this is differs from the respective answers in \cite{Li:2015rfa} by a factor of 2 in the $S$ and $T$ sectors since they use a different normalisation of the symmetric and anti-symmetric projectors. Also note that the $O(c^0)$ OPE coefficients are negative in certain sectors because of our normalisation of the conformal blocks.},     
\begin{equation}\label{MFTsolsonf}
\begin{split}
& C^{(S)}_{\Delta,\ell}=\frac{C^{MFT}_{n,\ell}}{N},\qquad C^{(T)}_{\Delta,\ell}=C^{MFT}_{n,\ell}, \qquad C^{(A)}_{\Delta,\ell}=-C^{MFT}_{n,\ell},\\
&
C^{MFT}_{n,\ell}=\frac{\pi  (\ell+1) 2^{-4 \Delta_\phi -2 l-4 n+7} (2 \Delta_\phi +\ell+2 n-2) \Gamma (n+\Delta_\phi -1) \Gamma (n+2 \Delta_\phi -3) \Gamma (\ell+n+\Delta_\phi )}{\Gamma (\Delta_\phi -1)^2 \Gamma (\Delta_\phi )^2 \Gamma (n+1) \Gamma (\ell+n+2) \Gamma \left(n+\Delta_\phi -\frac{3}{2}\right) \Gamma \left(\ell+n+\Delta_\phi -\frac{1}{2}\right)}\\& ~~~~~~~~~~~~~~~~\times \Gamma (\ell+n+2 \Delta_\phi -2).
\end{split}
\end{equation}
with $\Delta=2\Delta_{\phi}+2n+\ell$. We note that only even spins appear in singlet and traceless symmetric representations whereas only odd spins appear in antisymmetric representation. \footnote{In general dimensions, we can use the expansions listed in \cite{Hogervorst:2013sma} to get $C^{MFT}_{n,\ell}$ in general dimensions.}

\subsubsection*{Solution for $O\left(c^{-1}\right)$}

At order $O(c^{-1})$, we write down the equations which we have to solve to find anomalous dimension from \eqref{crossing2sc} and \eqref{confdesonf}. We will focus on the coefficient of the non-analytic pieces proportional to $\log z \log (1- \bar z)$ and restrict ourselves to finite spin cut-off $L$. 

\begin{eqnarray}\label{crossingksonf}
\alpha^\mu_{n,l}= M_{SO(N),f}^{\mu \nu} \tilde{\a}^\nu_{n,l},
\end{eqnarray}
where, 
\begin{eqnarray}
\a^\mu_{n,l}&=& \sum_n \sum^L_{\ell} C^{(\mu)}_{n,\ell} \gamma^{(\mu)}_{n,\ell} \frac{\bar{z}}{1-\bar{z}} \Bigg( z^{n+\ell} \bar{z}^{n-1} \, F_{\Delta_\phi+n+\ell}(z) \tilde{F}_{\Delta_{\phi}+n-1}(1-\bar{z})-(n+\ell)\leftrightarrow (n-1)\Bigg), \nonumber\\
\tilde{\a}^\mu_{n,l}&=& \sum_n \sum^L_{\ell} C^{(\mu)}_{n,\ell} \gamma^{(\mu)}_{n,\ell}\frac{z-1}{z}   \Bigg( (1-z)^{n+\ell} (1-\bar{z})^{n-1} F_{\Delta_{\phi}+n-1}(1-\bar{z}) \tilde{F}_{\Delta_{\phi}+n+\ell}(z)-(n+\ell)\leftrightarrow (n-1)\Bigg).\nonumber\\
\end{eqnarray}
and $\mu=1,2,3$ denotes $S,T$ and $A$ respectively and we have used $$F_\a(x)=\, _2F_1(\a,\a;2\a;x),\qquad \tilde{F}_\a(x)=-\frac{\Gamma(2\a)}{\Gamma(\a)^2}\, _2F_1(\a,\a;1;x).$$ We now project using orthogonality conditions of the Hypergeometric function,
\begin{eqnarray}\label{orthohyp}
\oint_C \frac{dz}{2\pi i} z^{m-m'-1} F_{\D + m} (z) F_{1-\D-m'}(z)=\delta_{m,m'},
\end{eqnarray}
to give us,
\begin{eqnarray}\label{crossingPsonf}
\b^\mu_{p,q}= M_{SO(N),f}^{\mu \nu} \b^\nu_{q,p}.
\end{eqnarray}
where, $\b^\mu_{p,q}$ is defined in equation \eqref{psonfdef}. The set of equations \eqref{crossingPsonf} can be solved by choosing particular values of $(p,q)$ analogous to \cite{Heemskerk:2009pn, Heemskerk:2010ty}. We have solved these set of equations upto a very large order of finite cut-off $L$ and present the pattern below. The number of undetermined anomalous dimensions can be encoded in the form of a partition function.
\begin{eqnarray}\label{cftpartsonf}
Z^{L=\text{even}}_{CFT, SO(N), f} = \frac{1}{8} (L+2) (3 L+4),\qquad
Z^{L=\text{odd}}_{CFT, SO(N), f} = \frac{1}{8} (L+1) (3 L+5).\nonumber\\
\end{eqnarray}
  
\subsection*{Adjoint scalars of $SO(N)$}

In this subsection we consider scalars charged under adjoint of $SO(N)$. The crossing equation is a bit more intricate and we follow the conventions of \cite{Li:2015rfa} and avoid giving explicit tedious expressions.  

\begin{equation}
\langle\phi_{i_1}^{i_2}(x_1)\phi_{j_1}^{j_2}(x_2)\phi_{k_1}^{k_2}(x_3)\phi_{l_1}^{l_2}(x_4)\rangle=\sum_r \left(t^r\right)^{i_2j_2k_2l_2}_{i_1i_2i_3i_4}\frac{G^{(r)}(u,v)}{x_{12}^{2\Delta_{\phi}}x_{34}^{2\Delta_{\phi}}},
\end{equation}

where the tensor structures $\left(t^r\right)^{i_2j_2k_2l_2}_{i_1i_2i_3i_4}$ have been listed in appendix B of \cite{Li:2015rfa} and we do not reproduce them here. The labels $r$ run over the irreducible representations that occur in the tensor product of two Adjoints of $SO(N)$, for convenience we label them by $(S, F, T, R, Ms, A)$ and they respectively correspond to the following irreducible representations of $SO(N)$.
$$\left({\mathbb I}, {\tyng(1,1)}, {\tyng(2)}, {\tyng(2,2)}, , {\tyng(2,1,1)}, {\tyng(1,1,1,1)}\right)$$

 The crossing equation then can be encoded in form of a matrix.

\begin{eqnarray}
G^{(\mu)}(u,v)= M^{\mu\nu}_{SO(N),a}\left( \frac{u}{v}\right)^{\Delta_\phi}G^{(\nu)}(u,v),
\end{eqnarray} 

where the matrix $M_{SO(N),a}$ is defined in \eqref{crossingmatsona}. The conformal block decomposition of $G^{(R)}(u,v)$ is given by,
\begin{eqnarray}
G^{(R)}(u,v) &=& \sum_{\Delta, \ell} C^{(R)}_{\Delta, \ell} g_{\D, \ell}(u,v).
\end{eqnarray} 
The spin support of the different sectors can be encoded as $(\text{even}, \text{odd}, \text{even}, \text{even}, \text{odd}, \text{even})$.  The projected crossing equations at $O\left( c^{-1}\right)$ are,
\begin{eqnarray}\label{crossingPsona}
\sigma^\mu_{p,q}= M_{SO(N),a}^{\mu \nu} \sigma^\nu_{q,p},
\end{eqnarray}
where $\sigma^\mu_{p,q}$ is defined in \eqref{Ysonadef}. We have solved these set of equations upto a very large order of $L$ and present the pattern below. The number of undetermined anomalous dimensions can be encoded in the form of a partition function.
\begin{eqnarray}\label{cftpartsona}
Z^{L=\text{even}}_{CFT, SO(N), a} = \frac{1}{4} (L+2) (3 L+4),\qquad
Z^{L=\text{odd}}_{CFT, SO(N), a} = \frac{1}{4} (L+1) (3 L+5).\nonumber\\
\end{eqnarray}
Note that these are exactly twice the partition function counting we evaluated for $SO(N)$ fundamental (see \eqref{cftpartsonf}).

\subsection*{Fundamental of $SO(4)$}\label{lvon}
In the previous subsections we have worked out the crossing equations for scalars charged under fundamental or adjoint of $SO(N)$ for a generic $N$. In general for low values of $N$, there are more tensor structures possible for the crossing equation (and similarly the bulk counting also is different). In this subsection we provide evidence that the correspondence holds true even for low values of $N$ by explicitly evaluating the spin support of crossing equations for $SO(4)$ fundamental scalars.
Let us consider the four point function of  identical scalars transforming under fundamental representation of $SO(4)$,
\begin{equation}
\langle\phi_i(x_1)\phi_j(x_2)\phi_k(x_3)\phi_\ell(x_4)\rangle=\frac{\mathcal{G}_{ijk\ell}(u,v)}{x_{12}^{2\Delta_{\phi}}x_{34}^{2\Delta_{\phi}}},
\end{equation}
Where $\mathcal{G}_{ijk\ell}(u,v)$ is modified from the large $N$ \cite{Cvitanovic:2008zz, Poland:2011ey, Kos:2015mba},
\begin{equation}\label{so4ce}
\begin{split}
\mathcal{G}_{ijk\ell}(u,v)&=\delta_{ij}\delta_{k\ell}G^{(S)}(u,v)+\left(\frac{\delta_{ik}\delta_{j\ell}+\delta_{i\ell}\delta_{jk}}{2}-\frac{1}{N}\delta_{ij}\delta_{k\ell}\right)G^{(T)}(u,v)+\left(\frac{\delta_{ik}\delta_{j\ell}-\delta_{i\ell}\delta_{jk}}{4}+\frac{\epsilon^{ijkl}}{2}\right)G^{(A)}(u,v)\nonumber\\
&+\left(\frac{\delta_{ik}\delta_{j\ell}-\delta_{i\ell}\delta_{jk}}{4}-\frac{\epsilon^{ijkl}}{2}\right)G^{(A')}(u,v).\nonumber\\
\end{split}
\end{equation}
Note the presence of the fully-antisymmetric $\epsilon^{ijkl}$ due to $SO(4)$. The crossing symmetry requires,
\begin{equation}
\mathcal{G}_{ijk\ell}(u,v)=\left(\frac{u}{v}\right)^{\Delta_{\phi}}\mathcal{G}_{kji\ell}(v,u).
\end{equation}
From this we arrive at the following constraints coming from crossing symmetry,

	\begin{equation}
	\begin{split}
	&G^{(S)}(u,v)=\left(\frac{u}{v}\right)^{\Delta_{\phi}} \Bigg(\frac{1}{4}G^{(S)}(v,u)+\frac{9}{16}G^{(T)}(v,u)+\frac{-3}{16}\left(G^{(A)}(v,u)+G^{(A')}(v,u)\right)\Bigg),\\
	& G^{(T)}(u,v)=\left(\frac{u}{v}\right)^{\Delta_{\phi}}\Bigg(G^{(S)}(v,u) +\frac{1}{4}G^{(T)}(v,u)+\frac{1}{4}\left(G^{(A)}(v,u)+G^{(A')}(v,u)\right)\Bigg),\\
	& G^{(A)}(u,v)=\left(\frac{u}{v}\right)^{\Delta_{\phi}}\Bigg(-G^{(S)}(v,u) +\frac{3}{4}G^{(T)}(v,u)+\frac{1}{4}\left(3G^{(A')}(v,u)-G^{(A)}(v,u)\right)\Bigg),\\
	& G^{(A')}(u,v)=\left(\frac{u}{v}\right)^{\Delta_{\phi}}\Bigg(-G^{(S)}(v,u) +\frac{3}{4}G^{(T)}(v,u)+\frac{1}{4}\left(3G^{(A)}(v,u)-G^{(A')}(v,u)\right)\Bigg).
	\end{split}
	\end{equation}



Now we have conformal block decomposition for each part of the correlator,
\begin{equation}
G^{(R)}(u,v)=\sum_{\Delta,\ell}C^R_{\Delta,\ell} ~ g_{\Delta,\ell}(u,v),
\end{equation}
where R could stand for singlet(S), traceless symmetric(T) or antisymmetric(A,A')  sectors. The double twist operators corresponding to the different irreducible sectors are as follows,
\begin{eqnarray}\label{dtso4f}
&&{\cal O}^{(S)}_{n,l}= \phi_i\square^{n}\partial^l\phi_i, \qquad {\cal O}^{(T)}_{n,l}= \phi_{(i}\square^{n}\partial^l\phi_{j)}- \frac{\delta_{ij}}{N}\phi_{k}\square^{2n}\partial^l\phi_{k}, \qquad , \qquad {\cal O}^{(A)}_{n,l}= \phi_{[i}\square^{n}\partial^l\phi_{j]} \nonumber\\
&& {\cal O}^{(A')}_{n,l}= \epsilon^{ijkm}\phi_{k}\square^{n}\partial^l\phi_{m} \nonumber\\
\end{eqnarray}
Note that the $S$ and $T$ sectors can take on only even values of spin and the $A, A'$ sectors have support only over odd spins. The MFT solutions are, 
\begin{equation}
\begin{split}
& C^{(S)}_{\Delta,\ell}=\frac{C^{MFT}_{n,\ell}}{4},\qquad C^{(T)}_{\Delta,\ell}=C^{MFT}_{n,\ell},\qquad  C^{(A)}_{\Delta,\ell}=-C^{MFT}_{n,\ell},\qquad C^{(A')}_{\Delta,\ell}=-C^{MFT}_{n,\ell}
\end{split}
\end{equation}
with $\Delta=2\Delta_{\phi}+2n+\ell$, $C^{MFT}_{n,\ell}$ is given by \eqref{MFTsolsonf} and only even spins appear in singlet and traceless symmetric representations whereas only odd spins appear in the two antisymmetric representations. The projected crossing equation at $O(c^{-1})$ is given by,
\begin{eqnarray}\label{crossingPso4f}
{\b'}^\mu_{p,q}= M_{SO(4),f}^{\mu \nu} {\b'}^\nu_{q,p}
\end{eqnarray}
where, ${\b'}^\mu_{p,q}$ is defined in equation \eqref{pso4fdef} and matrix  $M_{SO(4), f}$ is defined in equation \eqref{cmson4}. The partition function for the solutions takes the form
\begin{eqnarray}\label{cftpartso4f}
Z^{L=\text{even}}_{CFT, SO(4), f} =\left(\frac{L}{2}+1\right) L+1,\qquad
Z^{L=\text{odd}}_{CFT, SO(4), f} = 2 \left(\frac{L-1}{2}+1\right)^2.\nonumber\\
\end{eqnarray}



\subsection{Constraints from crossing: $SU(N)$}
In this section we consider scalars charged under the fundamental and anti-fundamental of $SU(N)$. We can consider the following correlator,
\begin{equation}\label{corr1}
\langle\phi_{i}(x_1)\phi^{\dagger j}(x_2)\phi_k(x_3)\phi^{\dagger\ell}(x_4)\rangle=\frac{\mathcal{G}^{j\ell}_{ik}(u,v)}{x_{12}^{2\Delta_{\phi}}x_{34}^{2\Delta_{\phi}}}=\sum_r \left(b^{(r)}\right)^{jl}_{ik}\frac{G^{(r)}(u,v)}{x_{12}^{2\Delta_{\phi}}x_{34}^{2\Delta_{\phi}}},
\end{equation}
where $\left(b^{(r)}\right)^{jl}_{ik}$ is given by,
\begin{equation}
\left(b^{(S)}\right)^{jl}_{ik}=\delta_i^j \delta_k^{\ell},\qquad \left(b^{(Adj)}\right)^{jl}_{ik}=\left(\delta_i^{\ell}\delta_k^j-\frac{1}{N}\delta_i^j \delta_k^{\ell} \right).
\end{equation}
Now the crossing symmetry requires (this is the equivalence of $s$ and $t$ channel),
\begin{equation}
\mathcal{G}^{j\ell}_{ik}(u,v)=\mathcal{G}^{j\ell}_{ki}(u,v).
\end{equation}
This gives us the following constraints,

\begin{equation}\label{crosssunf1}
\begin{split}
& G^{(S)}(u,v)-\frac{1}{N} G^{(Adj)}(u,v)=\left(\frac{u}{v}\right)^{\Delta_{\phi}}G^{(Adj)}(v,u),\\
& G^{(Adj)}(u,v)=\left(\frac{u}{v}\right)^{\Delta_{\phi}}\Bigg(G^{(S)}(v,u)-\frac{1}{N} G^{(Adj)}(v,u)\Bigg).
\end{split}
\end{equation}

In this crossing equation, the sum over spins for the $S$ and $Adj$ sector runs over all spins. To be more precise, let us label the sum over even and odd spins for a particular irreducible sector in the following manner \cite{Rattazzi:2010yc}
\begin{eqnarray}
\sum_{\ell= \text{even}} C^{(R)}_{\Delta,\ell} g_{\D, \ell}(u,v) = G^{(R)}_+, \qquad \sum_{\ell= \text{odd}} C^{(R)}_{\Delta,\ell} g_{\D, \ell}(u,v)  = G^{(R)}_-.
\end{eqnarray} 
Reflection positivity of the euclidean correlator \eqref{corr1} implies that, on both sides of the crossing equation \eqref{crosssunf1}, the sum over spins run over both even and odd spins with the same sign. In equations, 

\begin{equation}\label{crosssunf2}
\begin{split}
& \left(G^{(S)}_+(u,v)+G^{(S)}_-(u,v)\right)-\frac{1}{N} \left(G^{(Adj)}_+(u,v)+G^{(Adj)}_-(u,v)\right)=\left(\frac{u}{v}\right)^{\Delta_{\phi}}\left(G^{(Adj)}_+(v,u)+G^{(Adj)}_-(v,u)\right),\\
& \left(G^{(Adj)}_+(u,v)+G^{(Adj)}_-(u,v)\right)=\left(\frac{u}{v}\right)^{\Delta_{\phi}}\Bigg(\left(G^{(S)}_+(v,u)+G^{(S)}_-(v,u)\right)-\frac{1}{N}\left(G^{(Adj)}_+(v,u)+G^{(Adj)}_-(v,u)\right)\Bigg).
\end{split}
\end{equation}

We can now consider the $u$-channel crossing of correlator \eqref{corr1}. Instead of directly evaluating it, we can consider the t-channel and the s-channel expansion of the following transposed correlator,
\begin{equation} \label{corr2}
\langle\phi_i(x_1)\phi_j(x_2)\phi^{\dagger k}(x_3)\phi^{\dagger \ell}(x_4)\rangle
\end{equation}

The $s$ and $t$- channel expansions are, 
\begin{eqnarray} \label{stchfund}
(12)(34) &\equiv& \frac{1}{x_{12}^{2\Delta_{\phi}}x_{34}^{2\Delta_{\phi}}} \Bigg(\left(\delta_{i}^{k}\delta_j^{\ell}+\delta_i^{\ell}\delta_j^k\right) G^{(Sym)}(u,v)+\left(-\delta_{i}^{k}\delta_j^{\ell}+\delta_i^{\ell}\delta_j^k\right)G^{(Anti-sym)}(u,v)\Bigg),\nonumber\\
(14)(23)&\equiv&\frac{1}{x_{12}^{2\Delta_{\phi}}x_{34}^{2\Delta_{\phi}}} \left(\frac{u}{v}\right)^{\Delta_{\phi}}\Bigg(\delta_i^{\ell}\delta_j^k G^{(S)}(v,u)+\left(\delta_i^k\delta_j^{\ell}-\frac{1}{N}\delta_i^{\ell}\delta_j^k\right)G^{(Adj)}(v,u)\Bigg),
\end{eqnarray}

where the $G^{(Sym)}$ and $G^{{(Anti-sym)}}$ run over even and odd spins respectively. Equating different tensor structures on both sides of $s=t$ equation,  will lead us to following constraint equations,

	\begin{equation}\label{crosssunf3}
	\begin{split}
	& \left(G^{(S)}_+(u,v)-G^{(S)}_-(u,v)\right)-\frac{1}{N} \left(G^{(Adj)}_+(u,v)-G^{(Adj)}_-(u,v)\right)=\left(\frac{u}{v}\right)^{\Delta_{\phi}}\Bigg(G^{(Sym)}(v,u)+G^{(Anti-sym)}(v,u)\Bigg),\\
	& \left(G^{(Adj)}_+(u,v)-G^{(Adj)}_-(u,v)\right)=\left(\frac{u}{v}\right)^{\Delta_{\phi}} \Bigg(G^{(Sym)}(v,u)-G^{(Anti-sym)}(v,u)\Bigg).
	\end{split}
	\end{equation}

Note the change in sign of the odd spin sum in LHS of \eqref{crosssunf3} compared to \eqref{crosssunf2}. This is because the transposed correlator is no longer reflection positive in the $t$-channel\footnote{
See section 2.1 of \cite{Rattazzi:2010yc} for a group theoretic understanding of this.}. Using orthogonality conditions of the Hypergeometric function (\eqref{orthohyp}) we get,
\begin{eqnarray}\label{crossingPsunf}
\kappa^\mu_{p,q}= M_{SU(N),f}^{\mu \nu} \kappa^\nu_{q,p},\qquad \tau^\mu_{p,q}= \tilde{M}_{SU(N),f}^{\mu \nu} \Omega^\nu_{q,p}
\end{eqnarray}
where $\kappa^\mu_{p,q}, \tau^\mu_{p,q}$ and $\Omega^\mu_{p,q}$ are defined in \eqref{Ysunfdef}. The matrices  $M_{SU(N),f}$ and $\tilde{M}_{SU(N),f}$ are defined in equation \eqref{crossingmat1sunf} and \eqref{crossingmat2sunf}.  The number of undetermined anomalous dimensions can be encoded in the form of a partition function.
\begin{eqnarray}\label{cftpartsunf}
Z^{L=\text{even}}_{CFT, SU(N), f} = \frac{3}{4} L (L+2)+1, \qquad
Z^{L=\text{odd}}_{CFT, SU(N), f} =\frac{3}{4} (L+1)^2.\nonumber\\
\end{eqnarray}

\subsection*{Adjoint scalars of $SU(N)$}

In this subsection we consider scalars charged under adjoint of $SU(N)$ \cite{Li:2015rfa, Berkooz:2014yda}. 

\begin{equation}
\langle\phi_{i_1}^{i_2}(x_1)\phi_{j_1}^{j_2}(x_2)\phi_{k_1}^{k_2}(x_3)\phi_{l_1}^{l_2}(x_4)\rangle=\sum_r \left(s^r\right)^{i_2j_2k_2l_2}_{i_1i_2i_3i_4}\frac{G^{(r)}(u,v)}{x_{12}^{2\Delta_{\phi}}x_{34}^{2\Delta_{\phi}}},
\end{equation}

where the tensor structures $\left(s^r\right)^{i_2j_2k_2l_2}_{i_1i_2i_3i_4}$ have been listed in appendix B.3 of \cite{Li:2015rfa}. The labels $r$ run over the irreducible representations that occur in the tensor product of two adjoints of $SU(N)$, for convenience we label them by $(S, Adj_{-}, Adj_+, AS, AA, SS)$. The crossing equation then can be encoded in form of a matrix.

\begin{eqnarray}\label{crossingsuna}
G^{(\mu)}(u,v)= M^{\mu\nu}_{SU(N),a}\left( \frac{u}{v}\right)^{\Delta_\phi}G^{(\nu)}(u,v),
\end{eqnarray} 

where the matrix $M_{SU(N),a}$ is defined in \eqref{crossingmatsuna}. The conformal block decomposition of $G^{(R)}(u,v)$ is given by,
\begin{eqnarray}
G^{(R)}(u,v) &=& \sum_{\Delta, \ell} C^{(R)}_{\Delta, \ell} g_{\D, \ell}(u,v).
\end{eqnarray} 
The spin support of the different sectors can be encoded as $(\text{even}, \text{odd}, \text{even}, \text{odd}, \text{even}, \text{even})$. We now project using the orthogonality of the Hypergeometric function \eqref{orthohyp} to get,
\begin{eqnarray}\label{crossingPsuna}
\Lambda^\mu_{p,q}= M_{SO(N),a}^{\mu \nu} \Lambda^\nu_{q,p}
\end{eqnarray}
where $\Lambda^\mu_{p,q}$ is defined in \eqref{Ysunadef}. The matrix  $M_{SU(N), a}$ is given in equation \eqref{crossingmatsuna}. The number of undetermined anomalous dimensions can be encoded in the form of a partition function.
\begin{eqnarray}\label{cftpartsuna}
Z^{L=\text{even}}_{CFT, SU(N), a} = \frac{1}{4} (L+2) (3 L+4), \qquad Z^{L=\text{odd}}_{CFT, SU(N), a} = \frac{1}{4} (L+1) (3 L+5).\nonumber\\
\end{eqnarray}
Note that these are exactly the partition function counting we evaluated for $SO(N)$ adjoint (see \eqref{cftpartsona}).

\subsection{Counting flat space S-matrices}\label{cfssm}
\paragraph{}  In this section we follow \cite{Chowdhury:2019kaq, Chowdhury:2020ddc} to evaluate flat space S-matrices for scalars in $D\geq 4$ and determine the number of bulk contact terms with finite support over spin. Local Lagrangians are isomorphic to flat space s-matrices upto field re-definitions and equations of motion. We evaluate this by deriving an integral formula using plethystic techniques. Let us summarise the procedure in brief	. We construct the single letter partition function for the particle with internal symmetry label and impose equations of motion \cite{Aharony:2003sx, Sundborg:1999ue}. We evaluate the multi particle partition function by plethystic exponentiation and project the resulting group theoretic expression onto the singlets of the space-time symmetry and the internal symmetry. From general group theoretic arguments, it is known that the flat space S-matrices can be organised by their $S_3$ transformation properties. More precisely,  
\begin{equation} \label{partfnsssmt} 
Z_{ \text{S-matrix}}(x) = \sum_{J} x^{\Delta_J} 
Z_{{\bf R_J}}(x)
\end{equation} 
where $Z_{\bf R_J}(x)$ are listed in \eqref{partfnsss} and are the partition functions of irreducible representations ${\bf R_J}$ of $S_3$. More generally, from the partition function \eqref{partfnsssmt}, we can count how many linearly independent flat-space S-matrices are there at a particular derivative order. As an aside, in \cite{Chowdhury:2019kaq}, the authors obtained a nice mathematical structure of the S-matrices in terms of local and bare module generators \footnote{see section 2.4 of \cite{Chowdhury:2019kaq} for a self contained discussion on modules in the context of flat space S-matrices.}. For scalars the local and bare module generators are the same and \eqref{partfnsssmt} can be viewed as a partition function encoding the various local module generators with their respective $S_3$ transformation properties. In order to generate the S-matrices, this way of viewing the partition function will play a crucial role for finding the spin support of the local Lagrangians of a particular derivative order. We will also construct the explicit local Lagrangians (or the local modules) for the respective partition functions.

\subsection*{Scalar flat-space S matrices with internal symmetry}
Using AdS/CFT correspondence we can say that the boundary operator $\phi^{(R)}$ which transforms under an irreducible representation $R$ of some global symmetry group is dual to the field $\Phi^{(R)}$, which transforms under the same representation. 
In this section we first enumerate and construct the Lorentz scalars that can be built out of 
scalar fields $\Phi^{(R)}$ charged under 
\begin{itemize}
	\item Fundamental and adjoint of $SO(N)$.
	\item Fundamental-anti fundamental and adjoint of $SU(N)$.
\end{itemize}

The single letter partition function for scalars charged under some internal symmetry is a simple generalisation of the scalars with no internal symmetry; it is given by 
\begin{eqnarray}\label{scalar-single}
i_\ts(x,y,z)&=&{\rm Tr}\,\,x^{\Delta} y_i^{H_i}y_i^{z_i}= \chi_R(z)(1-x^2)\denom(x,y).\nonumber\\
\denom(x,y) &=&\Big(\prod_{i=1}^{D/2}(1-x y_i)(1-x y_i^{-1})\Big)^{-1}\qquad \qquad \qquad \qquad{\rm for \,\, D\,\, even},\nonumber \\
&=&\Big((1-x)\prod_{i=1}^{\lfloor D/2\rfloor}(1-x y_i)(1-x y_i^{-1})\Big)^{-1}\qquad \qquad \,\,{\rm for \,\, D\,\, odd}.
\end{eqnarray}
Here $H_i$ and $z_i$ stands for the Cartan elements of $SO(D)$ and $G$ respectively. $\denom(x,y)$ encodes the tower of derivatives on $\Phi(x)$ keeping track of the degree and the charges under the Cartan subgroup of $SO(D)$. The factor $\chi_R(z)$, basically the character of the representation $R$ of the internal symmetry group $G$, keeps track of the internal symmetry of the field $\Phi^{(R)}(G)$. For scalar fields charged under fundamental, adjoint representation of $SO(N)$ and adjoint representation of $SU(N)$, the Bose symmetrized multi letter partition function consisting of four letters is given by: 
\be\label{4-particle}
\begin{split}
	i_\ts^{(4)}(x,y,z)&=\frac{1}{24}\Big(i^4_\ts(x,y,z) +6 i^2_\ts(x,y,z) i_\ts(x^2,y^2,z^2)+3i^2_\ts(x^2,y^2,z^2)+8i_\ts(x,y,z)i_\ts(x^3,y^3,z^3)\\&+6i_\ts(x^4,y^4,z^4)\Big).
\end{split}
\ee  
Once we construct this, we recall that the equivalence class of scalar Lagrangians are given by scalar quartic polynomials (along with derivatives) modulo polynomials that are total derivatives. This is easily implemented by dividing the four letter partition function by   
$\denom(x,y)$, the generator for towers of derivatives. 
$$i_\ts^{(4)}(x,y,z)/\denom{(x,y)}$$
Finally to project onto the singlet sector of both $SO(D)$ and $SO(N)/SU(N)$, we perform a Haar integral over the Haar measure of the respective groups. Schematically this is given by,  
\be\label{singlet-proj}
I^R_\ts(x):=\oint  d\mu_{G}~ \oint  d\mu_{SO(D)}~  i_\ts^{(4)}(x,y,z)/\denom(x,y),
\ee  
where $d\mu_{SO(D)}$ is the haar measure associated with the Lorentz group $SO(D)$ and $d\mu_{G}$ is the haar measure associated with the colour group $G$. Using techniques outlined in appendix C of \cite{Chowdhury:2020ddc} and appendix H.1 of \cite{Chowdhury:2019kaq}, the Haar integral over the $SO(D)$ can be performed and  \eqref{singlet-proj} then takes the schematic form, 
\begin{eqnarray}\label{scalar-proj-colour}
I^R_\ts(x):= \oint d\mu_{G}~&\left(\frac{\chi^{G}_R(z^2) \chi^{G}_R(z)^2}{4 \left(1-x^4\right)}+\frac{\chi^{G}_R(z^4)}{4 \left(1-x^4\right)}+\frac{\chi^{G}_R(z)^4}{24 \left(x^2-1\right)^2}+\frac{\chi^{G}_R(z^2)^2}{8 \left(x^2-1\right)^2}\right.\nonumber\\
&\left. +\frac{\chi^{G}_R(z^3)\chi^{G}_R(z)}{3 \left(x^4+x^2+1\right)}\right).
\end{eqnarray}
We delegate the evaluation of the Haar colour integrals to the appendices (see appendix \ref{PI}) and present the results in the main section\footnote{See also \cite{Henning:2015daa, Henning:2017fpj, deMelloKoch:2017dgi, deMelloKoch:2018klm, Kobach:2018pie, Kobach:2017xkw, Melia:2020pzd} for recent progress using similar formalism and related interesting applications.}.

\begin{table}
	
\end{table}

\subsubsection*{$SO(N)$: fundamental and adjoint}

For scalars charged under the fundamental and the adjoint representation of $SO(N)$, using the integrals listed in table \ref{sofund}, \eqref{scalar-proj-colour} evaluates to the partition function, 

	\be \label{partfnsonfund}
	I^{f}_{\ts,~ SO(N)}(x) = \frac{1+x^2+x^4}{(1-x^4)(1-x^6)} =Z_{{\bf 3}}, \qquad 	I^{a}_{\ts,~ SO(N)}(x) = \frac{2+2x^2+2x^4}{(1-x^4)(1-x^6)} =2Z_{{\bf 3}}.
	\ee 
Note that the evaluation has been done for large $N$ \cite{Chowdhury:2020ddc}. From the analysis done in \cite{Chowdhury:2019kaq}, we see that this is the partition function corresponding to a local module that transforms in the ${\bf 3_S}$ of $S_3$. This is an reducible representation of $S_3$.
\begin{equation}
{\bf 3}={\bf 1_S} \oplus {\bf 2_M},
\end{equation}
where recall that ${\bf 1_S}$ and ${\bf 2_M}$ are the irreducible one and two dimensional representations of $S_3$. Let us suppose that the colour module generators transforming the ${\bf 3}$ of $S_3$ is denoted by $|e^{(1,2,3)}\rangle$. We follow the same conventions of \cite{Chowdhury:2019kaq}, where $|e^{(1)}\rangle, |e^{(2)}\rangle, |e^{(3)}\rangle$ are the module elements which are symmetric under particle swap $(3 \leftrightarrow 4)$, $(2 \leftrightarrow 4)$ and $(2 \leftrightarrow 3)$ respectively. 
 The local module at $2r$ order in derivatives is then given by the scalar products of the module generators with polynomial of mandelstam invariants which transform in the same irreducible representation. In equations, considering the basis of polynomials which transform in ${\bf 3}$ of $S_3$ to be given by $(f(t,u), f(t,s), f(s,u))$, one constructs the elements of the most general descendant module as 
$$\left(f(t,u)|e^{(1)}\rangle,~ f(t,s)|e^{(2)}\rangle,~ f(s,u)|e^{(3)}\rangle\right)$$
where $f(i,j)= i^{r-k}j^{k}+ i^{k}j^{r-k}$. The corresponding local scalar modules and its descendants can be easily obtained from the following tower of Lagrangians.

	\begin{eqnarray}\label{slagsonfund} 
	L_{SO(N),f} = \sum_{m,n}  a_{m,n} \prod_{b=1}^m\prod_{c=1}^n\left(\partial_{\m_b} \partial_{\n_c} \Phi_i \Phi_i \right)  \left(\partial^{\m_b} \Phi_j \partial^{\n_c} \Phi_j \right).
	\end{eqnarray}
	\begin{eqnarray}\label{slagsonadj}  
	L^1_{SO(N),a} &=& \sum_{m,n}  a_{m,n} \prod_{b=1}^m\prod_{c=1}^n \left(\partial_{\m_b} \partial_{\n_c} \Phi_{ij} \Phi_{ji} \right)  \left(\partial^{\m_b} \Phi_{kl} \partial^{\n_c} \Phi_{lk} \right),\qquad \nonumber\\
	L^2_{SO(N),a} &=& \sum_{m,n}  a_{m,n} \prod_{b=1}^m\prod_{c=1}^n \left(\partial_{\m_b} \partial_{\n_c} \Phi_{ij} \partial^{\m_b} \Phi_{jk}  \Phi_{kl} \partial^{\n_c} \Phi_{li} \right).
	\end{eqnarray}

Our condensed notation for the derivatives can be explained by considering the following expression,
\be\label{partfnsonadj}
\prod_{b=1}^m \,\,\partial_{\mu_b} \,\Phi_1\, \partial^{\mu_b} \,\Phi_2 \equiv \partial_{\mu_1}\partial_{\mu_2}\ldots \partial_{\mu_m} \,\Phi_1\,\,\partial^{\mu_1}\partial^{\mu_2}\ldots \partial^{\mu_m} \,\Phi_2.
\ee
for some operators $\Phi_1$ and $\Phi_2$. The same notation is also used for the second tower of derivatives indexed as $\partial_{\nu_c}$. In particular, each term denotes a Lorentz invariant Lagrangian term with $2m+2n$ derivatives. 
The Lagrangians \eqref{slagsonfund} and \eqref{slagsonadj} encode the most general higher derivative Lagrangians that we can build out of identical scalars charged under $SO(N)$  fundamental and adjoint respectively. The partition functions \eqref{partfnsonfund} tells us of the $S_3$ transformation properties of the local Lagrangian structures.  


\subsubsection*{$SU(N)$: adjoint} 

From \cite{Chowdhury:2020ddc} we can compute the large $N$ plethystic integrals for $SU(N)$. Using table \ref{suadj}, \eqref{scalar-proj-colour} for $SU(N)$ adjoint representation turns out to be  

	\be \label{partfnsunadj}
	I^{a}_{\ts,~ SU(N)}(x) = \frac{2+2x^2+2x^4}{(1-x^4)(1-x^6)}=2Z_{{\bf 3}}.
	\ee 
The local modules transform in ${\bf 3_S}$ and the associated Lagrangians are given by

	\begin{eqnarray}\label{slagsunadj}  
	L^1_{SU(N),a} &=& \sum_{m,n}  a_{m,n} \prod_{b=1}^m\prod_{c=1}^n \left(\partial_{\m_b} \partial_{\n_c} \Phi^i_j \Phi^j_i \right)  \left(\partial^{\m_b} \Phi^k_l \partial^{\n_c} \Phi^l_k \right), \nonumber\\
	L^2_{SU(N),a} &=& \sum_{m,n}  a_{m,n} \prod_{b=1}^m\prod_{c=1}^n\left(\partial_{\m_b} \partial_{\n_c} \Phi^i_j \partial^{\m_b} \Phi^j_k  \Phi^k_l \partial^{\n_c} \Phi^l_i \right). \nonumber\\
		\end{eqnarray}

\subsubsection*{$SU(N)$: fundamental}                     
We now turn to the problem of evaluating flat space S-matrices of scalars charged under fundamental and anti-fundamental representation of $SU(N)$. The four letter partition function relevant for counting singlets is a bit different for this case than \eqref{4-particle}. Two of the scalar fields is charged under the fundamental representation while the other two have to be charged under the anti-fundamental representation for non-zero singlets. The symmetry group is now $\Z_2 \otimes \Z_2$ instead of the full $S_4$ while the $S_3$ is replaced by $\Z_2$.\footnote{We thank Abhijit Gadde for discussions regarding this point.} The singlet condition is therefore given by 
\bea
Z_{f-\bar{f}}&=& \sum_{i_1, i_2, j_3, j_4}\langle i_1 i_2 j_3 j_4|y^I y^J \frac{\left(1+P_{12}P_{34}\right)}{2} | i_1 i_2 j_3 j_4\rangle=\frac{1}{2}\left( \rho(x)^2\bar{\rho}(x)^2 + \rho(x^2)\bar{\rho}(x^2) \right)\nonumber\\
&=& S^2(\rho \otimes \bar{\rho})
\eea 	



where $\rho$ denotes the fundamental representation and $\bar{\rho}$ denotes the anti-fundamental representation of $SU(N)$. The resulting modules are charged under $\Z_2$ of the $\Z_2 \times \Z_2$ and the number of such modules are given by

\bea\label{chargedmodules}
Z^\pm_{f-\bar{f}}&=& \sum_{i_1, i_2, j_3, j_4}\langle i_1 i_2 j_3 j_4|y^I y^J \left( \frac{1 \pm P_{12}}{2}\right)\frac{\left(1+P_{12}P_{34}\right)}{2} | i_1 i_2 j_3 j_4\rangle= \begin{cases}
	S^2(\rho) \otimes S^2(\bar{\rho}),\\
	\Lambda^2(\rho) \otimes \Lambda^2(\bar{\rho})\\
\end{cases}
\eea 	
For $N\geq 3$, we find $S^2(\rho) \otimes S^2(\bar{\rho})=\Lambda^2(\rho) \otimes \Lambda^2(\bar{\rho})=1$. Without loss of generality, we can take particles 1 and 4 to transform in the fundamental representation while particles 2 and 3 transform in the anti-fundamental representation. The two $\Z_2$ invariant modules for $N\geq 3$ \footnote{We have used \cite{lieart} to evaluate \eqref{chargedmodules} for various values of $N$.} are 

\bea
|e_1\rangle= (\phi_i(p_1)\phi^i(p_2))(\phi_j(p_4)\phi^j(p_3)), \qquad |e_2 \rangle= (\phi_i(p_1)\phi^i(p_3))(\phi_j(p_4)\phi^j(p_2)).
\eea 
Defining the modules under $P_{34}$/$P_{12}$ symmetry and anti symmetry as 
\bea
|e^+\rangle=|e_1 \rangle+|e_2 \rangle, \qquad |e^-\rangle=|e_1 \rangle-|e_2 \rangle,
\eea 
the most general descendant module is generated by 
\bea\label{sunfgenerators}
J_1&=&\left( \sum_{ m,n} a_{m,n} (stu)^m (s^2+t^2+u^2)^n \right) \left(|e^+\rangle\right),\nonumber\\
J_2&=&\left( \sum_{ m,n} a_{m,n} (stu)^m (s^2+t^2+u^2)^n \right) 
\bigg( (s +t)|e^+\rangle \bigg),\nonumber\\
J_3&=&\left( \sum_{ m,n} a_{m,n} (stu)^m (s^2+t^2+u^2)^n \right) 
\bigg( (s -t)|e^-\rangle \bigg),\nonumber\\
J_4&=&\left( \sum_{ m,n} a_{m,n} (stu)^m (s^2+t^2+u^2)^n \right) 
\bigg( \left( s^2+ t^2-2u^2 \right) |e^+\rangle \bigg),\nonumber\\
J_5&=&\left( \sum_{ m,n} a_{m,n} (stu)^m (s^2+t^2+u^2)^n \right) 
\bigg( \left( s^2- t^2\right) |e^-\rangle \bigg),\nonumber\\
J_6&=&\left( \sum_{ m,n} a_{m,n} (stu)^m (s^2+t^2+u^2)^n \right) 
\bigg( \left( s^2t- t^2s+t^2u-u^2t+u^2s-s^2u\right) |e^-\rangle \bigg).
\eea


The local Lagrangian which gives rise to this module is given by

	\be\label{slagsunfund}
	L_{SU(N),f} = \sum_{m,n}  a_{m,n} \prod_{b=1}^m\prod_{c=1}^n \left(\partial_{\m_b} \partial_{\n_c} \Phi_i \partial^{\n_b}\bar{\Phi}^i \right)  \left( \Phi_j  \partial^{\m_b}\bar{\Phi}^j \right). 
	\ee 	

The generators \eqref{sunfgenerators} are in one-to-one correspondence with the descendent module from the Lagrangian \eqref{slagsunfund}, where the factors of $s$ and $t$ count two derivative orders each. The number of linearly independent S-matrices at a given order of $2k$ derivatives are the number of solutions to $6m+4n=2k+\alpha_i$ for each $J_i$ (to be precise, $\a_1=0, \a_2=\a_3=2, \a_4=\a_5=4, \a_6=6$). The spin support for the module generated by this Lagrangian is the same as that of a module transforming in ${\bf 6}= {\bf 1_S} + 2~ {\bf 2_M} + {\bf 1_A}$ of $S_3$.

\subsection{Counting independent data labelled by spin}
\paragraph{} In this section we evaluate the number of distinct contact terms contributing to a spin $l$ exchange ( following \cite{Heemskerk:2009pn}). Consider, as an warm-up, S-matrices generated by the scalars without global symmetry, that can be put into the generic form \cite{Chowdhury:2019kaq} 
\be
(s t u)^m(st +tu+us)^n |e_{\bf S}\rangle.
\ee 
In order to count for spin exchanged, we note that  $s, t$ and $u$ in centre-of-mass frame can be expressed as,
\be
t= \frac{-s}{2}\left( 1-\cos \theta \right), \qquad u= \frac{-s}{2}\left( 1+\cos \theta \right).
\ee                              
Consider a scattering process where the highest spin being exchanged is $L$ (an even integer since we are considering identical scalars) corresponding to a non-coloured scalar S-matrix of 2k derivative order. We can write the following set of equations,   
\be 
3m+2n=k, \qquad 2m+2n=L
\ee 

Thus we can see that the allowed terms contributing to highest spin exchange $L$ are derivative terms with $k: L, L+1, \cdots L+ \frac{L}{2}$. Total number of flat space s-matrices therefore contributing upto $L$ exchange is \cite{Heemskerk:2009pn},
\be\label{symmlsc}
\sum_{a=0}^{\frac{L}{2}}(1+a) = \frac{(L+2)(L+4)}{8}
\ee 
Recall that a most general S-matrix is given by a $S_4$ invariant polynomial of momenta and global symmetry charges (and also polarizations for spinning particles). The $\Z_2 \times \Z_2$ group is a normal subgroup for $S_4$ and consequently the S-matrices can be labelled by their $S_3$ transformation properties alone. As explained in the previous section, we can view the most general S-matrix as being generated by the scalar product of polynomials of mandelstam variables with the module. S-matrix from the local Lagrangians listed in the previous sections thus are given by a linear combination of the $S_3$ modules listed in \cite{Chowdhury:2019kaq},

\begin{eqnarray}\label{mjilmo}
I_1&=&(s t u)^m(st +tu+us)^n \left( |e_{\bf S}\rangle \right),\nonumber\\
I_2&=&\left( \sum_{ m,n} a_{m,n} (stu)^m (s^2+t^2+u^2)^n \right) 
\bigg( (s +t)|e_{\bf 2_M}^{(1)}\rangle +(t +u)|e_{\bf 2_M}^{(2)}\rangle + (u+s)|e_{\bf 2_M}^{(3)} \rangle \bigg),\nonumber\\
I_3&=&\left( \sum_{ m,n} a_{m,n} (stu)^m (s^2+t^2+u^2)^n \right) 
\bigg( \left( s^2+ t^2-2u^2 \right) |e_{\bf 2_M}^{(1)}\rangle + 
\left( t^2+u^2-2s^2 \right) |e_{\bf 2_M}^{(2)}\rangle +\nonumber\\
&&\left( u^2+ s^2-2t^2 \right)|e_{\bf 2_M}^{(3)} \rangle \bigg),\nonumber\\
I_4&=&\left( \sum_{ m,n} a_{m,n} (stu)^m (s^2+t^2+u^2)^n \right) 
\bigg( \left( s^2t-t^2s+t^2u-u^2t+u^2s-s^2u \right) |e_{\bf 1_A}\rangle\bigg).
\end{eqnarray} 
where $|e_{\bf S}\rangle , |e_{\bf 2_M}^{(i)} \rangle$ and $|e_{\bf 1_A}\rangle$ are the module for Lagrangians transforming in $\bf{1_S}$, $\bf{2_M}$ and $\bf{1_A}$ respectively (these are the only three irreducible representations of $S_3$). In the context of present paper, the local modules are basically contractions of generator matrices under which the fields are charged. Note that $I_1, I_2, I_3$ and $I_4$ essentially encode the information contained in the partition functions \eqref{partfnsonfund},\eqref{partfnsonadj} and \eqref{partfnsonadj}. We have to find the corresponding number of flat-space S-matrices contributing to a particular spin $L_0$ from each of $I_1$, $I_2$, $I_3$ and $I_4$ respectively and sum them up. The contribution from $I_1$ has been already worked out in \eqref{symmlsc} \cite{Heemskerk:2009pn}. The contribution from $I_2, I_3$ and $I_4$ can be worked out similarly and is given by,

\begin{eqnarray}\label{spnsupp}
n_{I_1}(L)&=&\frac{1}{2} \left(\left\lfloor \frac{L}{2}\right\rfloor +1\right) \left(\left\lfloor \frac{L}{2}\right\rfloor +2\right),\qquad
n_{I_2} (L) = \frac{1}{2} \left(\left\lfloor \frac{L-1}{2}\right\rfloor +1\right) \left(\left\lfloor \frac{L-1}{2}\right\rfloor +2\right), \nonumber\\
n_{I_3} (L) &=& \frac{1}{2} \left(\left\lfloor \frac{L-2}{2}\right\rfloor +1\right) \left(\left\lfloor \frac{L-2}{2}\right\rfloor +2\right),\qquad
n_{I_4} (L) = \frac{1}{2} \left(\left\lfloor \frac{L-3}{2}\right\rfloor +1\right) \left(\left\lfloor \frac{L-3}{2}\right\rfloor +2\right),\nonumber\\
\end{eqnarray} 
where $\lfloor x \rfloor$ implies the integer less than or equal to $x$. Therefore the maximum number of linearly independent flat space S-matrices contributing upto spin $L_0$ exchange for a module that is transforming in a ${\bf 6}$ of $S_3$ is given by 
\begin{eqnarray}
n(L_0) = n_{I_1}(L_0)+2n_{I_2}(L_0)+2n_{I_3}(L_0)+ n_{I_4}(L_0)
\end{eqnarray}
We list the spin support of the scalar S-matrices eqn \eqref{slagsonfund} \eqref{slagsonadj} \eqref{slagsunadj} and \eqref{slagsunfund} in order of their $S_3$ representations in table \ref{spinsupport}.

\begin{table}[h!]
	\begin{center}
		\begin{tabular}{|c|c|c|} 
			\hline
			\textbf{S-matrix Lagrangian} & \textbf{$S_3$ representations} & \textbf{Spin support $L$}\\
			\hline
			$L_{SO(N),f}$& ${\bf 3}$ & $n_{I_1}(L)+n_{I_2}(L)+n_{I_3}(L)$ \\
			\hline
			$L^1_{SO(N),a}$& ${\bf 3}$ & $n_{I_1}(L)+n_{I_2}(L)+n_{I_3}(L)$ \\
			\hline
			$L^2_{SO(N),a}$& ${\bf 3}$ & $n_{I_1}(L)+n_{I_2}(L)+n_{I_3}(L)$ \\
			\hline
			$L^1_{SU(N),a}$& ${\bf 3}$ & $n_{I_1}(L)+n_{I_2}(L)+n_{I_3}(L)$ \\
			\hline
			$L^2_{SU(N),a}$& ${\bf 3}$ & $n_{I_1}(L)+n_{I_2}(L)+n_{I_3}(L)$ \\
			\hline
			$L_{SU(N),f}$& ${\bf 6}$ & $n_{I_1}(L)+2n_{I_2}(L)+2n_{I_3}(L)+ n_{I_4}(L)$ \\
			\hline
		\end{tabular}
		\caption{Spin support for scalar S-matrices}\label{spinsupport}
	\end{center}
\end{table}

We obtain perfect agreement with the respective partition functions for spin support from CFT computation (see \eqref{cftpartsonf},\eqref{cftpartsona}, \eqref{cftpartsunf} and \eqref{cftpartsuna}). 
\subsubsection*{Counting bulk Lagrangians and support on spin for $SO(4)$}
The multi particle partition function for $SO(4)$ gives the following result,
	\be \label{partfnso4fund}
I^{f}_{\ts,~ SO(4)}(x) = \frac{1+x^2+x^4+x^6}{(1-x^4)(1-x^6)} =Z_{{\bf 3}}+ Z_{{\bf 1_A}}.
\ee 
This implies in addition to the usual Lagrangians \eqref{slagsonfund}, we have  the additional Lagrangian given by the bulk Lagrangian,
\begin{eqnarray}\label{slagso4fund} 
L_{SO(4),f} = \sum_{m,n}  a_{m,n} \prod_{b=1}^m\prod_{c=1}^n \epsilon^{ijkl} \left(\partial_{\m_b} \partial_{\n_c} \phi_i \phi_j\partial^{\m_b} \phi_k\partial^{\n_c} \phi_l \right).
\end{eqnarray}
The module generators transform in a ${\bf 1_A}$ of $S_3$ and are given by, 
\begin{eqnarray}
&&\left( \sum_{ m,n} a_{m,n} (stu)^m (s^2+t^2+u^2)^n \right) 
\bigg( \left( s^2t-t^2s+t^2u-u^2t+u^2s-s^2u \right) |e_{\bf 1_A}\rangle\bigg),\nonumber\\
&&|e_{\bf 1_A}\rangle = \epsilon^{ijkl}.
\end{eqnarray}
The spin support is therefore given by,
\begin{eqnarray}
	n_{SO(4),f} &=& n_{I_1}(L)+n_{I_2}(L)+n_{I_3}(L)+ n_{I_4}(L),
\end{eqnarray}
where $n(I_i)$s are given in \eqref{spnsupp} and we obtain perfect agreement with \eqref{cftpartso4f}.

\subsection{Counting Bulk ambiguities using Functionals}\label{cbauf}
The conformal bootstrap program in position space has been very successful numerically, though it's often hard to analytically constrain the OPE data. Recently interest has been rekindled in analytic constrains in the form of pursuit of extremal analytic functionals \cite{Mazac:2016qev}, where it was found for CFTs in $d=1$ and also for special external dimensions of the operators. It was further extended to general cases in \cite{Mazac:2018mdx, Mazac:2018ycv, Mazac:2018qmi, Kaviraj:2018tfd, Paulos:2019gtx, Paulos:2020zxx, Mazac:2018biw} in one dimension. These constructions naturally lead to a crossing symmetric formulation in terms of all three channels unlike the more conventional bootstrap equation. These $d=1$ functionals have nice positivity properties and that enabled the authors to construct the extremal functionals analytically and put analytic bounds on OPE coefficient. 

Construction of analytic functionals in higher dimension was first carried out in \cite{Mazac:2019shk} but this time the crossing symmetry was respected with respect to two channels unlike $d=1$ case. These functionals (without any further non trivial modifications) don't have good positivity properties. Nevertheless they are still useful and it has many advantages when applied to holographic CFTs and perturbative CFTs (\cite{Penedones:2019tng, Carmi:2020ekr}). In \cite{Mazac:2019shk} the authors proposed the following expansion of any unitary Regge superbounded \footnote{to be defined below.} CFTs,
\begin{equation}
\mathcal{G}(z,\bar{z})=\mathcal{G}^s+\mathcal{G}^t,
\end{equation}
where $\mathcal{G}^s$ and $\mathcal{G}^t$ are such that,
\begin{equation}
dDisc_t(\mathcal{G}^s)=0,\,\,\,\,\,\,dDisc_s(\mathcal{G}^t)=0,
\end{equation}
where $dDisc$ means double discontinuity of the correlator. Using this property $\mathcal{G}^s$ and $\mathcal{G}^t$ can be expanded in t channel and s channel double trace conformal blocks (denoted by $G^t_{\Delta_{n,\ell},\ell}(z, \bar{z})$ and $G^s_{\Delta_{n,\ell},\ell}(z, \bar{z})$ respectively),
\begin{equation}
 \mathcal{G}(z,\bar{z})=\sum_{n,\ell} \alpha^s_{n,\ell} G^s_{\Delta_{n,\ell},\ell}(z, \bar{z})+\beta^s_{n,\ell} \partial_{\Delta}G^s_{\Delta_{n,\ell},\ell}(z, \bar{z})+(s\leftrightarrow t).
 \end{equation} 
$\alpha^s_{n,\ell} ,\beta^s_{n,\ell} ,\alpha^t_{n,\ell},\beta^t_{n,\ell}  $ form a dual basis of linear functionals and by definition $G^t(z,\bar{z})=G^s(1-z,1-\bar{z})$. When these elements of dual basis act on crossing equation they lead to sum rule of the form,
\begin{equation}
\sum_{\Delta,\ell}a_{\Delta,\ell} \omega[F^{\Delta_{\phi}}(z,\bar{z})]=0,\qquad F^{\Delta_{\phi}}(z,\bar{z})= \left(z \bar{z}\right)^{-\Delta_{\phi}}G^s_{\Delta, \ell}(z, \bar{z}) -\left((1-z)(1-\bar{z})\right)^{-\Delta_{\phi}}G^t_{\Delta, \ell}(z, \bar{z})
\end{equation}
where $\omega$ can be $\alpha$ or $\beta$. These are dual to GFF solutions, i.e., they have double zeroes at $\Delta_{n,\ell}=2\Delta_{\phi}+2n+\ell$. Therefore in a perturbative expansion around GFF (e.g. holographic CFTs) the double trace operators get suppressed and the sum rules constrain the single trace data.

In this context we should discuss the u channel regge limit to introduce few terminology. The u channel regge limit is defined by taking the limit $z,\bar z$ goes to $i \infty$ with $\frac{z}{\bar{z}}$ fixed. The correlators $\mathcal{G}(z,\bar{z})$ are bounded by $(z\bar{z})^{\frac{J-1}{2}}$. The correlators are superbounded if the Regge spin is negative, on the other hand the unitary correlators are only bounded as $J<2$. Note that $\mathcal{G}^s$ and $\mathcal{G}^t$ are linearly independent when they are inside the space of superbounded functions. But in the space of bounded functions these are not independent. The dual basis satisfy the following orthonormality conditions,
 \begin{equation}
 \begin{split}
 \alpha^q_{n,\ell}[G^r_{\Delta_{n',\ell'},\ell}]=\delta^{q r}\delta_{n n'}\delta_{\ell \ell'},\qquad \alpha^q_{n,\ell}[\partial_{\Delta}G^r_{\Delta_{n',\ell'},\ell}]=0,\\
 \beta^q_{n,\ell}[G^r_{\Delta_{n',\ell'},\ell}]=0,\qquad  \beta^q_{n,\ell}[\partial_{\Delta}G^r_{\Delta_{n',\ell'},\ell}]=\delta^{q r}\delta_{n n'}\delta_{\ell \ell'}.
 \end{split}
 \end{equation}
 where $q, r$ stand for either $s$ or $t$ channel.

Now  we use the techniques described in \cite{Mazac:2019shk}  to write down functionals which can act on the crossing equation we have written down above for various internal symmetry groups. Though it's an straightforward generalization but the class of functionals are different and therefore, we find worth mentioning few aspects of it here. This method is democratic to spacetime dimensions unlike HPPS functionals which were applied to two and four spacetime dimensions.  The functional action on the conformal blocks has a nice physical interpretation in terms of exchange Witten diagram and a class of contact diagrams which are bounded in the $u$ channel Regge limit, eg, a Regge bounded exchange Witten diagram has the following OPE decomposition in the direct channel,
\begin{equation}
'\tilde{W}^s_{\Delta,\ell}(z,\bar{z})=G^s_{\Delta,\ell}(z, \bar{z})+\sum_{n,\ell} A_{n,\ell} G^s_{2\Delta_{\phi}+2n+\ell,\ell}(z, \bar{z})+\sum_{n,\ell} B_{n,\ell} \partial G^s_{2\Delta_{\phi}+2n+\ell,\ell}(z, \bar{z}).
\end{equation}

$A_{n,\ell}$ and $B_{n,\ell}$ are related to functional actions on block as,
\begin{equation}
A_{n,\ell}=-\alpha^s_{n,\ell}\,\,\,\,\,\, B_{n,\ell}=-\beta^s_{n,\ell}.
\end{equation}

The functional actions on the block also has a nice integral representation,
\begin{equation}
\beta_{n,\ell}=\int [dw][d\bar{w}]\mathcal{H}(w,\bar{w}) G^s_{\Delta,\ell}(w,\bar{w}),
\end{equation}
where $\mathcal{H}(w,\bar{w})$ is constrained by the $u-$ channel Regge growth of the correlator and $[dw]=\frac{dw}{2\pi i}$. In \cite{Mazac:2019shk}  a recipe to construct such kernel was given when the correlator is Regge superbounded, so the following fall of kernel was sufficient,
\begin{equation}
\mathcal{H}(w,\bar{w})\rightarrow w^{-1}, \,\,\,\,\,\, \text{as}\,\, w\rightarrow \infty, \bar{w}\rightarrow \infty, \text{with}\,\, w/\bar{w}=\text{fixed}.
\end{equation}

But in a unitary theory the $u-$ channel Regge spin is bounded by 2. So we require kernels to have stronger fall off as we probe the Regge limit. In \cite{Caron-Huot:2020adz} was shown that we can get these better behaved kernels by taking suitable linear combinations of those poor behaved kernels. These kernels were acted upon a $s-t$ antisymmetric crossing equation, i.e.
\begin{equation}
\beta_{n,\ell}:F^{\Delta_{\phi}}(z,\bar{z})=\bigg (\left(z \bar{z}\right)^{-\Delta_{\phi}}G^s(z,\bar{z})-\left((1-z)(1-\bar{z})\right)^{-\Delta_{\phi}} G^t(z,\bar{z})\bigg),
\end{equation}
Further the above action can be written down as an action only on direct channel blocks such as,
\begin{equation}
\begin{split}
\beta_{n,\ell}& =\int [dw][d\bar{w}]\mathcal{H}(w,\bar{w}) \bigg (\left(w \bar{w}\right)^{-\Delta_{\phi}}G^s(w,\bar{w})-\left((1-w)(1-\bar{w})\right)^{-\Delta_{\phi}} G^t(w,\bar{w})\bigg)\\
& =\int [dw][d\bar{w}] \left(\mathcal{H}(w,\bar{w})-\mathcal{H}(1-w,1-\bar{w})\right)\left(w \bar{w}\right)^{-\Delta_{\phi}}G^s(w,\bar{w}).
\end{split}
\end{equation}

For unitary theories we expect that,
\begin{equation}
\mathcal{H}(w,\bar{w})-\mathcal{H}(1-w,1-\bar{w})\rightarrow w^{-3}, \,\,\,\,\,\, \text{as}\,\, w\rightarrow \infty, \bar{w}\rightarrow \infty, \text{with}\,\, w/\bar{w}=\text{fixed}.
\end{equation}

These are also called spin-2 convergent functionals. Examples of such functionals include the following \cite{Caron-Huot:2020adz} (see appendix \ref{relnbetwnbb}),
\begin{equation}\label{nu}
\nu_{i,j}=(i+1)^2\hat{\beta}_{i+1,j}-(j+1)^2\hat{\beta}_{i,j+1}-(i-j)(i+j+1)\hat{\beta}_{i,j}.
\end{equation}

These are not good functionals for the crossing equations we have at hand as these are not $s-t$ symmetric. Indeed their action on the crossing symmetric function $F^{\Delta_{\phi}}(z,\bar{z})$ is non trivial but their action on crossing antisymmetric combination $H^{\Delta_{\phi}}(z,\bar{z})$ \footnote{$H^{\Delta_{\phi}}(z,\bar{z})=\bigg (\left(z \bar{z}\right)^{-\Delta_{\phi}}G^s(z,\bar{z})+\left((1-z)(1-\bar{z})\right)^{-\Delta_{\phi}} G^t(z,\bar{z})\bigg)$.} is trivially zero. In general in the analysis of crossing symmetry equations of four scalar fields with global symmetry $F^{\Delta_{\phi}}(z,\bar{z})$ and $H^{\Delta_{\phi}}(z,\bar{z})$ both appear. One way forward is that we can take a larger set of Regge bounded functionals so that the final kernels arising from their subtractions will have a better fall off, i.e. instead of demanding better fall off of the combination of $\mathcal{H}(w,\bar{w})-\mathcal{H}(1-w,1-\bar{w})$, individually we can improve it. \footnote{Our equations have different sectors corresponding to irreducible representations of the internal symmetry group. So one can also consider subtracting equations arising from different sectors to achieve better fall off of the kernel. We do not try this here. } Let us explicitly write down few examples of such functionals,
\begin{equation}
\begin{split}
&\tilde{\nu}_{1,2}=\nu_{1,2}+2\nu_{0,2}-3\nu_{0,3}, \qquad \tilde{\nu}_{1,3}=\nu_{1,3}+6\nu_{0,3}-8\nu_{0,4},\\
& \tilde{\nu}_{1,4}=\nu_{1,4}+12\nu_{0,4}-15\nu_{0,5},\qquad  \tilde{\nu}_{2,3}=-3\nu_{2,3}-3\nu_{1,3}+4\nu_{1,4},\\
& \tilde{\nu}_{2,4}=8\nu_{2,4}+20\nu_{1,4}-25\nu_{1,5}.
\end{split}
\end{equation}
All of these combinations are well behaved in the $u-$ channel Regge limit (to be precise their fall off property in individual channels is $\frac{1}{w^3}$ in contrast with \eqref{nu}, which has similar fall-off behaviour only in the combination $s-t$.) so that we can act them on our $O(N)$ crossing equations,
\begin{equation}
\begin{split}
& \left(z\bar{z}\right)^{-\Delta_{\phi}}\mathcal{G}^{S(s)}(z,\bar{z})-\left((1-z)(1-\bar{z})\right)^{-\Delta_{\phi}} \bigg(\frac{1}{N}\mathcal{G}^{S(t)}(z,\bar{z})+\frac{(N+2)(N+1)}{2N^2} \mathcal{G}^{T(t)}(z,\bar{z})\\
&+\frac{1-N}{2N}\mathcal{G}^{A(t)}(z,\bar{z}\bigg)=0,\\
& \left(z\bar{z}\right)^{-\Delta_{\phi}}\mathcal{G}^{T(s)}(z,\bar{z})-\left((1-z)(1-\bar{z})\right)^{-\Delta_{\phi}} \left(\mathcal{G}^{S(t)}(z,\bar{z})+\frac{(N-2)}{2N} \mathcal{G}^{T(t)}(z,\bar{z})+\frac{1}{2}\mathcal{G}^{A(t)}(z,\bar{z}\right)=0,\\
& \left(z\bar{z}\right)^{-\Delta_{\phi}}\mathcal{G}^{A(s)}(z,\bar{z})-\left((1-z)(1-\bar{z})\right)^{-\Delta_{\phi}} \left(-\mathcal{G}^{S(t)}(z,\bar{z})+\frac{(N+2)}{2N} \mathcal{G}^{T(t)}(z,\bar{z})+\frac{1}{2}\mathcal{G}^{A(t)}(z,\bar{z}\right)=0.
\end{split}
\end{equation}
So we can act $\tilde{\nu}$ on the above equations and those will give us nonperturbative sum rules for $O(N)$ fundamental theories. We can act these to crossing equations arising from other groups and different representations as well without any further modifications.

Let us now turn our attention to AdS contact diagrams with four scalar fields transforming in the fundamental representation of $O(N)$. These contact terms have no dDisc and therefore they can be expanded in the $s-$ channel conformal blocks and its derivatives with dimensions $\Delta=2\Delta_{\phi}+2n+\ell$,
\begin{equation}
\begin{split}
& \mathcal{G}^S(z,\bar{z})=\sum_{n,\ell} \left(C^{S,1}_{n,\ell}+\frac{1}{2}C^0_{n,\ell} \gamma^S_{n,\ell}\partial_{\Delta}\right)G_{2\Delta_{\phi}+2n+\ell,\ell},\qquad \mathcal{G}^T(z,\bar{z})=\sum_{n,\ell} \left(C^{T,1}_{n,\ell}+\frac{1}{2}C^0_{n,\ell} \gamma^T_{n,\ell}\partial_{\Delta}\right)G_{2\Delta_{\phi}+2n+\ell,\ell},\\
& \mathcal{G}^A(z,\bar{z})=\sum_{n,\ell} \left(C^{A,1}_{n,\ell}+\frac{1}{2}C^0_{n,\ell} \gamma^A_{n,\ell}\partial_{\Delta}\right)G_{2\Delta_{\phi}+2n+\ell,\ell}.
\end{split}
\end{equation}

Also there will be only even spin exchanges in the singlet and traceless symmetric sector, whereas there will be only odd spin exchanges in the antisymmetric sector. Now we can act our $\tilde{\nu}$ on these equations to find the following relation in $d=4$ and $\Delta_{\phi}=2$,
\begin{equation}
\begin{split}
& \gamma^S_{2,0}=-\frac{27}{7}\left(\gamma^S_{1,0}-\gamma^S_{0,0}\right),\qquad \gamma^T_{2,0}=-\frac{27}{7}\left(\gamma^T_{1,0}-\gamma^T_{0,0}\right),\qquad \gamma^A_{1,1}=\frac{27}{35}\gamma^A_{0,1}.
\end{split}
\end{equation}

These results agree with our previous computation. Also we know that we have two contact terms, one zero derivative and a two derivative term whose Regge behaviour is bounded by spin 2. Both terms contribute to anomalous dimensions of singlet and traceless symmetric sector and only the two derivative term contributes to anomalous dimension of antisymmetric operator. We have constructed more functionals and checked that they all agree with results derived from other method whenever available. Note that in this computation that we have presented, although we have ``assumed" spin support, in principle, following \cite{Caron-Huot:2020adz}, one can find linear combinations of $\tilde{\nu}_{i,j}$ functionals which tells us that the spin support is finite. To give further example the following functional falls off as $w^{-4}$,
\begin{equation}
\begin{split}
& \tilde{\nu}_{1,4}-\frac{3}{2}\tilde{\nu}_{2,3}-\frac{3}{4}\tilde{\nu}_{1,3}\\
&=-\frac{3}{4}\left(6 \nu_{0,3}-24 \nu_{0,4}+20\nu_{0,5}+3\nu_{1,3}-4\nu_{1,4}+2\nu_{2,3}\right)
\end{split}
\end{equation}
and it can bootstrap contact diagrams in AdS which has support till spin 2. This way we can always start with the kernels which are well behaved in the regge limit and then find a combination of them to improve the Regge behaviour further. We have not exhausted the algorithm but hopefully have been able to convey to the interested reader, the novelty of this approach.  

\section{Majorana fermions in $d=1$ with global symmetry}\label{mfi1dwgs}

In this section we consider solutions to crossing in the case when external particles are majorana fermions in $d=1$. We  will use the functional techniques to derive the CFT data. The structure of $D=2$ S-matrices will be related to bootstrap solutions of $d=1$ CFTs. 

\subsection{Counting using functionals}
In this case the functionals are defined as an integral action on the crossing equations but they lead to crossing symmetric functionals unlike the cases we described in subsection \ref{cbauf}, where three channel crossing symmetry is broken. The important difference between these two constructions are as follows: first, there are spinning exchange operators in higher dimensions and there are only scalars or fermions to consider in one dimensional CFTs because of lack of rotation in one dimension. More significant difference comes from the fact that the functionals in higher dimension acts on channels individually and therefore its action on the crossing equation is trivial. This lead to a two channel crossing symmetric construction of functionals in higher dimensions. On the other hand in one dimension we will see that the kernels are built in a way so that its action on the crossing equation is non-trivial and that leads to three channel crossing symmetric functionals having important positivity properties which are lacking in those functionals discussed in previous sections unless we take infinite combinations of them cleverly \cite{Caron-Huot:2020adz}.

To be precise, let us write the conformal block expansion of four point function of majorana fermions charged under fundamental of $SO(N)$. 

\begin{equation}
\begin{split}
\left\langle \psi_i(x_1) \psi_j(x_2) \psi_k(x_3) \psi_l (x_4)\right\rangle=& \frac{1}{x_{12}^{2\Delta_{\psi}}x_{34}^{2\Delta_{\psi}}}\bigg(\delta_{ij}\delta_{kl} \sum_{\Delta}a^S_{\Delta}G_{\Delta}(z) +(\delta_{il}\delta_{jk}+\delta_{ik}\delta_{jl}-\frac{2}{N}\delta_{ij}\delta_{kl}) \sum_{\Delta}a^T_{\Delta}G_{\Delta}(z) +\\
& (\delta_{il}\delta_{jk}-\delta_{ik}\delta_{jl})\sum_{\Delta}a^A_{\Delta}G_{\Delta}(z) \bigg),
\end{split}
\end{equation}
where,
\begin{equation} \nonumber
z^2=\frac{x_{12}^2x_{34}^2}{x_{13}^2x_{24}^2}\,,\,\,\,\,\,x_{ij}=(x_i-x_j).
\end{equation}
This can be expressed more compactly in the following manner
\begin{equation}\label{crosssonff}
\sum_{\Delta} C^S_{\Delta} \begin{pmatrix}
0\\
F_{\Delta}^{\Delta_{\psi}}(z) \\
H_{\Delta}^{\Delta_{\psi}}(z)
\end{pmatrix} + \sum_{\Delta} C^T_{\Delta} \begin{pmatrix}
F_{\Delta}^{\Delta_{\psi}}(z)\\
(1-\frac{2}{N})F_{\Delta}^{\Delta_{\psi}}(z) \\
-(1+\frac{2}{N})H_{\Delta}^{\Delta_{\psi}}(z)
\end{pmatrix} +\sum_{\Delta} C^A_{\Delta} \begin{pmatrix}
-F_{\Delta}^{\Delta_{\psi}}(z)\\
F_{\Delta}^{\Delta_{\psi}}(z) \\
-H_{\Delta}^{\Delta_{\psi}}(z)
\end{pmatrix}=0,
\end{equation}
where $F_{\Delta}^{\Delta_{\psi}}(z)= z^{-2\Delta_{\psi}} G_\Delta(z) - (1-z)^{-2\Delta_{\psi}} G_\Delta(1-z)$, $H_{\Delta}^{\Delta_{\psi}}(z)= z^{-2\Delta_{\psi}} G_\Delta(z) + (1-z)^{-2\Delta_{\psi}} G_\Delta(1-z)$  and $G_\Delta(z)$ is the $d=1$ conformal block. 
$$G_\Delta (z)= z^{\Delta} \, _2F_1(\Delta,\Delta;2\Delta;z). $$

The functional action can be represented by,  

\begin{equation}
\sum_{r,\Delta} a^r_{\Delta} (\omega\cdot\mathcal{F}^r_{\Delta})=0,
\end{equation}
where the functional action itself is represented by the following integral action.
\begin{align}
\omega\cdot\mathcal{F}^r_\Delta=&-\int_{\frac{1}{2}}^{\frac{1}{2}+i \infty} dz \,  \ \{ f_1(z),f_2(z),f_3(z) \}\cdot \mathcal{F}^r_{\Delta}  \ + \ \int_{\frac{1}{2}}^1 dz \, \ \{ g_1(z),g_2(z),g_3(z) \}\cdot \mathcal{F}^r_{\Delta} \,. \nonumber\\
\end{align}

$\mathcal{F}^r_\Delta$ for $O(N)$ group is defined as the column vectors in \eqref{crosssonff} and $r$ denotes the representations $(S,T,A)$. To illustrate this let us consider the simplified kernels for $\Delta_{\psi}=\frac{1}{2}$ \cite{Ghosh:2021ruh}\footnote{Kernels for general $\Delta_{\psi}$ can be found in \cite{Ghosh:2021ruh} and the subtraction scheme described above does not depend on dimension of external operators. So we quoted a simplified example for $\Delta_{\psi}=\frac{1}{2}$ in our discussion above.},
\begin{equation}
\begin{split}
\vec{f^S_m}(z)=& \{a_m \frac{2}{N}\bigg(\frac{P_{1+2m}(\frac{z-2}{z})}{z}+\frac{P_{1+2m}(\frac{1+z}{z-1})}{1-z}\bigg), a_m \frac{N+1}{N}\bigg(\frac{P_{1+2m}(\frac{z-2}{z})}{z}-\frac{P_{1+2m}(\frac{1+z}{z-1})}{1-z}\bigg),\\
& a_m \frac{N-1}{N}\bigg(\frac{P_{1+2m}(\frac{z-2}{z})}{z}-\frac{P_{1+2m}(\frac{1+z}{z-1})}{1-z}\bigg)\}\\
\vec{f^T_m}(z)=& \big\{-a_m \bigg(\frac{P_{1+2m}(\frac{z-2}{z})}{z}+\frac{P_{1+2m}(\frac{1+z}{z-1})}{1-z}\bigg),-a_m \bigg(\frac{P_{1+2m}(\frac{z-2}{z})}{z}+\frac{P_{1+2m}(\frac{1+z}{z-1})}{1-z}\bigg), \\
& a_m \bigg(\frac{P_{1+2m}(\frac{z-2}{z})}{z}-\frac{P_{1+2m}(\frac{1+z}{z-1})}{1-z}\bigg)\big\}\\
\vec{f^A_m}(z)=& \big\{b_m \bigg(\frac{P_{2m}(\frac{z-2}{z})}{z}+\frac{P_{2m}(\frac{1+z}{z-1})}{1-z}\bigg),-\frac{b_m}{2} \bigg(\frac{P_{2m}(\frac{z-2}{z})}{z}+\frac{P_{2m}(\frac{1+z}{z-1})}{1-z}\bigg), \\
& \frac{b_m}{2} \bigg(\frac{P_{2m}(\frac{z-2}{z})}{z}-\frac{P_{2m}(\frac{1+z}{z-1})}{1-z}\bigg)\big\}
\end{split}
\end{equation}

where $P_{m}(z)$ are Legendre polynomials and 
\begin{equation}
a_m=-\frac{\Gamma^2(2+2m)}{\pi^2\Gamma(3+4m)}.\,\,\,\,\,\,\,\, b_m=-\frac{\Gamma^2(1+2m)}{\pi^2\Gamma^2(1+4m)}
\end{equation}

Now in the Regge limit, the kernel has a fall off $O(\frac{1}{z^2})$ in Singlet and Traceless symmetric sector whereas it has slower fall off, $O(\frac{1}{z})$,  in the Antisymmetric channel. We know that the unitary  CFT correlator grows at most like a constant in the Regge limit. So the third component of the kernel requires to be improved by further subtraction. In presence of global symmetry, there is exactly one contact term (deformation) in $AdS_2$ which is regge bounded, i.e., the four fermi interaction term without any derivatives, which vanishes due to anticommuting property if there is no color. So we have to subtract among the unimproved kernels for different $m$ such that the subtracted kernels will have correct fall off to bootstrap the regge bounded unitary CFT correlators. One such choice is to subtract $m=0$ functional and our improved kernel will take the following form,
\begin{equation}
\tilde{\vec{f^{r}_m}}(z)=\vec{f^{r}_m}(z) +c_m \vec{f^A_0}(z),
\end{equation}

with $c_m$ is determined by demanding $O(\frac{1}{z^2})$ fall off in all sectors and this depends on $\Delta_{\psi}$ and $\Delta$. Thus the idea is that to bound correlation functions that are badly regge behaved, we subtract the prefunctionals amongst each other. The form of the above kernels are fixed such that the following orthogonality conditions are satisfied,
\begin{equation}
\begin{aligned}
\alpha_n^{\mf r}(\mf s,\Delta_m^{\mf s})&=\delta_{n,m} \delta^{\mf r \mf s}\,,& \qquad \partial_{\Delta}  \alpha_n^{\mf r}(\mf s,\Delta_m^{\mf s})&=-d_n^{\mf r, \mf s}\delta_{m,0},\\
\beta_n^{\mf r}(\mf s,\Delta_m^{\mf s})&=0\,,& \qquad \partial_{\Delta}  \beta_n^{\mf r}(\mf s,\Delta_m^{\mf s})&=\delta_{n,m} \delta^{\mf r \mf s}-c_n^{\mf r, \mf s}\delta_{m,0},
\end{aligned}
\end{equation}

where $c^{\mf r,\mf s}_n$ are some constants which depends on the subtractions we have to make such that the integration is finite. Also the labels $\mf r,\mf s$ stands for different irreps of $O(N)$. The double trace operator dimensions for different sectors is given by,  $\Delta_m^{S/T}=2\Delta_{\psi}+2m+1$  and $\Delta_m^{ A}=2\Delta_{\psi}+2m$. Then if we have the CFT correlator which grows like $z^p$, we have to demand stronger fall off for the kernel and that will introduce more subtractions. These are in one-to-one correspondence number of bulk contact terms upto $2p$ derivatives. E.g. for $p=1$,
 we can take combinations like,
 \begin{equation}
 \tilde{\vec{f^{r}_m}}(z)=\vec{f^{r}_m}(z) + c_m \vec{f^{A}_0}(z)+d_m \vec{f^{A}_1}(z)  +e_m \vec{f^{S}_0}(z),
 \end{equation}
so that this falls off like $\frac{1}{z^3}$ in all sectors. This tells us that there are three contact terms if we consider contact terms involving at most with two derivatives. Proceeding in a similar manner, this exercise tells us that as we increase $4$ derivatives, another $3$ contact terms are added to the list. The counting problem of the number of subtractions for a particular Regge behaviour can be encoded in the form of a partition function,

\begin{eqnarray}
I^{SO(N)}_{\tf, f}(x) &=& \frac{1+2x^2}{(1-x^4)}.
\end{eqnarray}   
This partition function is to be understood as a series expansion about $x=0$. Sum of coefficient upto $x^n$ in this expansion denotes the number of subtractions required from our basis in order to bootstrap a correlator which grows like $x^n$ in the Regge limit. We will show that this matches an independent counting of flat space S-matrix in the next section .

\subsection{Majorana fermion flat space S-matrices in 1+1 dimensions}\label{mffssi11d}  \paragraph{} We first enumerate and construct the Lorentz scalars that can be built out of massive majorana fields in two spacetime dimensions charged under 
fundamental of $SO(N)$. Consider a theory of massive majorana $\psi_\alpha^{R}(G)$ charged under some irreducible representation $R$ of an internal symmetry group $G$ ($\alpha$ is the spinor index which we will not explicitly indicate from this point onwards). We wish to study the most general local action for this theory, retaining only those terms that affect four fermion scattering. We consider the equation of motion for our field $\psi^{R}(G)$ to be 
\begin{eqnarray}\label{majeqn}
\slashed{\partial} \psi^{R}(G)= m \psi^{R}(G), \qquad -\slashed{\partial} \bar{\psi}^{R}(G)= m \bar{\psi}^{R}(G)
\end{eqnarray}
We adopt the majorana conventions of \cite{freedman2012supergravity} for our gamma matrices. 
\begin{eqnarray}
\gamma^0=\begin{pmatrix}
0 & 1 \\
-1 & 0\\
\end{pmatrix} = i \sigma_2, \qquad \gamma^1=\begin{pmatrix}
0 & 1 \\
1 & 0\\
\end{pmatrix} = \sigma_1
\end{eqnarray}  
In this representation the majorana condition and the majorana conjugate becomes, 
\begin{equation}\label{majoranacondition}
\psi^\star= \psi, \qquad \bar{\psi}= \psi^T C  
\end{equation}
where $C= i \gamma^0$. In Appendix \ref{majscat}, we construct the explicit plane wave solutions to equation \eqref{majeqn}, necessary to construct the flat space S-matrices. For majorana fermion fields we first construct the multi letter partition function consisting of four letters by Bose anti-symmetrizing the single letter partition function.  The four-letter partition function - relevant for counting quartic Lagrangians is given by: 
\be\label{4-particle fermion}
\begin{split}
	i_\tf^{(4)}(x,y,z)&=\frac{1}{24}\Big(i^4_\tf(x,y,z) -6 i^2_\tf(x,y,z) i_\tf(x^2,y^2,z^2)+3i^2_\tf(x^2,y^2,z^2)+8i_\tf(x,y,z)i_\tf(x^3,y^3,z^3)\\&-6i_\tf(x^4,y^4,z^4)\Big).
\end{split}
\ee  
where $i_\tf(x,y,z)$ is the majorana single letter partition function where $x$ keeps track of the operator dimension while $y_i$ and $z_i$ are chemical potentials corresponding to the cartan charges of the Lorentz group and the internal symmetry group respectively. Once we construct this, we recall that the equivalence class of fermion Lagrangians are given by fermion quartic polynomials (along with derivatives) modulo polynomials that are total derivatives. This is easily implemented by dividing the four letter partition function by   
$\denom(x,y)$, the generator for towers of derivatives. 
$$i_\tf^{(4)}(x,y,z)/\denom{(x,y)}$$
Finally to project onto the singlet sector of both space time symmetry group and internal symmetry, we perform a Haar integral over the Haar measure of the respective groups. Schematically this is given by,  
\be\label{singlet-proj-ferm}
I^R_\tf(x):=\oint  d\mu_{G}~ \oint  d\mu_{SO(D)}~  i_\tf^{(4)}(x,y,z)/\denom(x,y).
\ee  
where $d\mu_{SO(D)}$ is the haar measure associated with the Lorentz group $SO(D)$ and $d\mu_{G}$ is the haar measure associated with the colour group $G$. In this subsection we will restrict the Lorentz group to $SO(2)$. The integral at hand, \eqref{singlet-proj-ferm}, therefore has two Haar integrals one of which pertains to projecting onto Lorentz singlets, while the other is to project onto the colour singlets. We perform the Haar integral for both using numerical techniques techniques used in \cite{Chowdhury:2019kaq}. 
\subsubsection{Fermions without colour}  

In this subsection we derive the partition function for Majorana fermions without any colour. The single letter partition function for majorana fermions is given by \cite{Dolan:2005wy, Aharony:2003sx}, 
\begin{eqnarray}\label{fermion-single}
i_\tf(x,y)&=&{\rm Tr}\,\,x^{\Delta} y_i^{H_i}= \chi_f(y)(1-x)\denom_2(x,y).\nonumber\\
\denom_2(x,y) &=&\Big((1-x y_1)(1-x y_1^{-1})\Big)^{-1}, \nonumber\\
\chi_f(y) &=& \left(\sqrt{y_1}+ \frac{1}{\sqrt{y_1}} \right)
\end{eqnarray}
Here $H_i$ stands for the Cartan elements of $SO(2)$. The denominator factor $\denom_2(x,y)$ encodes the tower of derivatives on $\psi(x)$ keeping track of the degree and the charges under the Cartan subgroup of $SO(2)$ while the factor $\chi_f(z)$, basically encodes the character of the spinor representation. These are necessary since we will eventually project onto singlets of $SO(2)$\footnote{Note that we recover the majorana fermion letter partition function in \cite{Aharony:2003sx} (see eqn B.11) if we set $y_i=0$ in \eqref{fermion-single}}. Although we have expressed this for $D=2$, in principle this can be extended to higher dimensions. Using numerical integration, we find that \eqref{4-particle fermion} evaluates to
	\begin{eqnarray}\label{fermionpleth}
	I_{\tf}(x) &=& \frac{x^2}{(1-x^4)}={\bf Z_A}.
	\end{eqnarray}
In this partition function, $x$ keeps track of the derivative order and indicates that the module transforms in the antisymmetric irreducible representation of $S_2$ (see appendix \ref{rtosami}). The number of independent fermion Lagrangians, at a particular derivative order $m$, is obtained from \eqref{fermionpleth} by taylor expanding this partition function about $x=0$ and looking at the coefficient of $x^m$. The Lagrangian giving rise to the flat space S-matrices can be listed as

	\begin{equation}\label{fermlagnocol}
	L^F=	\sum_{m,n}  a_{m,n} \prod_{b=1}^m\prod_{c=1}^n(\partial_{\m_b} \partial_{\n_c}\bar{\psi}\partial^{\m_b}\psi)( \partial^{\n_c}\bar{\psi}\psi),
	\end{equation}
where we have used the following condensed notation for the derivatives,
\be
\prod_{b=1}^m \,\,\partial_{\mu_b} \,{\cal O}_1\, \partial^{\mu_b} \,{\cal O}_2 \equiv \partial_{\mu_1}\partial_{\mu_2}\ldots \partial_{\mu_m} \,{\cal O}_1\,\,\partial^{\mu_1}\partial^{\mu_2}\ldots \partial^{\mu_m} \,{\cal O}_2.
\ee
for some operators ${\cal O}_1$ and ${\cal O}_2$. The same notation is also used for the second tower of derivatives indexed as $\partial_{\nu_c}$. In particular, each term denotes a Lorentz invariant Lagrangian term with $2m+2n$ derivatives. Now the Lagrangians, as written, are not linearly independent. The linearly independent grassmann modules are as follows.
\begin{itemize}
	\item $m+n=even\geq 0$: There is no linearly independent S-matrix. These S-matrices are given by $m^\alpha$ times the lower derivative-order S-matrices. \\
	
	\item $m+n=\text{odd}\geq1$: There is one linearly independent module which is given by  any $a_{m,n}$.\\
\end{itemize}


\subsubsection{Fermions charged under fundamental of $SO(N)$}
The single letter partition function for majorana fermions charged under fundamental of $SO(N)$ is a simple generalisation of \eqref{fermion-single} and is given by \cite{Dolan:2005wy, Aharony:2003sx}, 
\begin{eqnarray}\label{fermion-single-colour}
i_\tf(x,y,z)&=&{\rm Tr}\,\,x^{\Delta} y_i^{H_i} y_i^{z_i}= \chi^{SO(N)}_{f}(z) \chi_f(y)(1-x)\denom_2(x,y).\nonumber\\
\chi^{SO(N)}_f( y) &=& \sum_{i=1}^{\lfloor N/2\rfloor}\left( y_i + \frac{1}{y_i} \right)\qquad \qquad \text{for N even},\nonumber\\
\chi^{SO(N)}_f( y) &=& \sum_{i=1}^{\lfloor N/2\rfloor}\left( y_i + \frac{1}{y_i} \right)+1\qquad ~\text{for N odd}.\nonumber\\
\end{eqnarray}
The Haar integrals for the colour is done using the large $N$ integrals listed in table \ref{sofund}, while the space time integral has been done numerically. The final result is given by,

\begin{eqnarray}\label{fermionplethcolour}
I^{SO(N)}_{\tf, f}(x) &=& \frac{1+2x^2}{(1-x^4)}={\bf Z_S}+2{\bf Z_A}.
\end{eqnarray}   

The Lagrangian which saturates the paritition function counting is given by 

	\begin{eqnarray}\label{fermlagsonfund}
	L^F_{SO(N),f}	=\sum_{m,n}  a_{m,n} \prod_{b=1}^m\prod_{c=1}^n(\partial_{\m_b} \partial_{\n_c}\bar{\psi}_i \partial^{\m_b}\psi_j)(\partial^{\n_c}\bar{\psi}_j\psi_i).\nonumber\\
	\end{eqnarray}

The linearly independent grassmann modules are as follows.
\begin{itemize}
	\item $m+n=even\geq 0$: There is one linearly independent S-matrix for any $a_{m,n}$.  \\
	
	\item $m+n=\text{odd}\geq1$: There are two linearly independent modules which are given canonically by $a_{m+n,0}$ and $a_{0, m+n}$.\\
\end{itemize}

\section{Conclusions}

In this paper we have revisited to the investigation of locality of bulk physics in AdS by a counting argument on both sides of the duality following \cite{Heemskerk:2009pn}. We considered CFTs with scalars in $d=4$ and fermions in $d=1$ charged under various global symmetry group. Then assuming a large central charge expansion we have counted the number of independent solutions to crossing equation at first non trivial order in $O\left(c^{-1}\right)$ expansion using HPPS functionals and also the analytic functionals introduced in \cite{Mazac:2019shk, Caron-Huot:2020adz, Ghosh:2021ruh}. The analytical functionals can be used to find CFT data in any spacetime dimension. There is a correspondence between number of independent CFT solutions and $S$ matrices in flat space which we evaluated through plethystic counting and obtained a perfect agreement. Apart from charged scalars we have also computed the $S$ matrices of charged fermions in two dimensions.

In \cite{Chowdhury:2019kaq, Chowdhury:2020ddc} the flat space four graviton and four gluon $S$ matrices were constructed. We would like to extend the notion of bulk locality in the sense of HPPS to spinning operators also. In the same spirit, it will be to interesting to consider three dimensional majorana fermions in the CFT \cite{Iliesiu:2015qra, upcoming}. The group theoretic counting of fermion S-matrices, introduced in this paper, can be generalised to $D=4$. We are hoping to extend the analytic functional methods of \cite{Mazac:2019shk, Caron-Huot:2020adz} to obtain the CFT data. This exercise of classification of flat space S-matrix counting also has implications for the S-matrix bootstrap \cite{Hebbar:2020ukp}.  We expect functional methods of \cite{Caron-Huot:2021enk} (as well as formalism of \cite{Kundu:2021qpi}) can be extended in presence of global symmetry and non-integer spin, putting the conjecture of \cite{Heemskerk:2009pn} on further concrete footing. Recently in \cite{Poland:2021xjs} the blocks for five point correlator was found. Therefore we can also test the notion of bulk locality in higher point functions given the plethystic counting for scalars are already known in literature. Another possibility is to push this computations to next order in large central charge expansion. With the CFT data found in this work we can now construct the thermal two point functions which are also dual to the same theories of charged fields as considered here in AdS but with non zero temperature. Following \cite{Iliesiu:2018fao, Alday:2020eua} we can assume that there are no new operators in the OPE of two fields apart from those which appeared in our analysis at zero temperature. Then using KMS conditions and assuming polynomial boundedness of thermal two point functions in the  Regge limit we can compute the correction to mean field theory thermal two point functions. In particular it will be interesting to explore this possibility for $d=3$ fermionic CFTs which will be dual to fields in $AdS_4$.  

Another interesting direction is to explore the Colour kinematics and Double Copy relations in AdS \cite{Armstrong:2020woi, Albayrak:2020fyp, Alday:2021odx, Zhou:2021gnu}. In \cite{Broedel:2012rc} it was pointed out that there is a tension in establishing CK duality in four point amplitudes due to EFT corrections to pure non abelian gauge field. It will be interesting to explore the status of the same in AdS for gluons, gravitons and fermion EFT corrections.

We leave a detailed analysis for further work.

\section*{Acknowledgements}
We would like to thank Aneesh P. B, A. Gadde, S. Hegde, A. Laddha and  M. Raman  for discussions. We would also like to thank A. Kaviraj and A. Sinha for comments on the manuscript. 
The work of SDC is supported by the Infosys Endowment for the study of the Quantum Structure of Spacetime. We would all also like to acknowledge our debt to the people of India for their steady support to the study of the basic sciences.

\appendix


\section{Details of crossing matrices}\label{docm}
In this section we write down explicitly the crossing matrices that we will refer to in the main text. The crossing matrix for scalars charged under fundamental of $SO(N)$ is given by, 

\begin{eqnarray}\label{cmsonf}
M_{SO(N),f}=&\begin{pmatrix}
 \frac{1}{N}& \frac{(N+2)(N-1)}{2N^2}& \frac{1-N}{2N}\\
1 &\frac{(N-2)}{2N} & \frac{1}{2}\\
-1 &\frac{(N+2)}{2N}& \frac{1}{2}
\end{pmatrix}
\end{eqnarray}

The matrix $\b^\mu_{p,q}$ used in \eqref{crossingPsonf} is defined as 
\begin{eqnarray}\label{psonfdef}
\b^S_{p,q}&=& \left(\sum^L_{\ell=0, \text{even}}\frac{\gamma'_S(-\ell+p-1,\ell) J^{\Delta -1}(-\ell+p-1,q)-\gamma'_S(p,\ell) J^{\Delta -1}(\ell+p+1,q)}{n}) \right)\nonumber\\
\b^T_{p,q}&=& \left(\sum^L_{\ell=0, \text{even}}\frac{\gamma'_T(-\ell+p-1,l) J^{\Delta -1}(-\ell+p-1,q)-\gamma'_T(p,\ell) J^{\Delta -1}(\ell+p+1,q)}{n}) \right)\nonumber\\
\b^A_{p,q}&=& -\left(\sum^L_{\ell=1, \text{odd}}\frac{\gamma'_A(-\ell+p-1,l) J^{\Delta -1}(-\ell+p-1,q)-\gamma'_A(p,\ell) J^{\Delta -1}(\ell+p+1,q)}{n}) \right)\nonumber\\
\end{eqnarray}

where
\begin{eqnarray}
\gamma'_i&=& \frac{2(\ell+1)(2\D+2n+\ell-2)}{(\D-1)^2}\gamma^{(i)}(n,\ell)\nonumber\\
\end{eqnarray}
and $J^{\Delta -1}(p,q)$ has been defined in \eqref{jdefn}. Note that the negative sign in $P^A_{p,q}$ is due to the negative sign in the solution of the MFT OPE coefficient in \eqref{MFTsolsonf}.

The crossing matrix for $SO(N)$ adjoint is given by \cite{Li:2015rfa},
\begin{eqnarray}\label{crossingmatsona}
M_{SO(N),a}=\left(
\begin{array}{cccccc}
\frac{2}{(\text{N}-1) \text{N}} & \frac{4 (\text{N}-2)}{(\text{N}-1) \text{N}} & \frac{4 (\text{N}-2) (\text{N}+2)}{(\text{N}-1) \text{N}^2} & \frac{\text{N}^3-7 \text{N}-6}{(\text{N}-1)^2 \text{N}} & \frac{-\text{N}^2+\text{N}+6}{\text{N}-\text{N}^2} & \frac{(\text{N}-3) (\text{N}-2)}{(\text{N}-1) \text{N}} \\
\frac{1}{2 (\text{N}-2)} & \frac{1}{2} & \frac{-\text{N}^2+2 \text{N}+8}{4 \text{N}-2 \text{N}^2} & \frac{-\text{N}^3+7 \text{N}+6}{4 (\text{N}-2)^2 (\text{N}-1)} & 0 & \frac{\text{N}-3}{2 (\text{N}-2)} \\
\frac{1}{2 (\text{N}-2)} & \frac{\text{N}-4}{2 (\text{N}-2)} & \frac{\text{N}^2-8}{2 (\text{N}-2) \text{N}} & \frac{\text{N}^3-6 \text{N}^2+5 \text{N}+12}{4 (\text{N}-2)^2 (\text{N}-1)} & \frac{3-\text{N}}{(\text{N}-2)^2} & -\frac{\text{N}-3}{2 (\text{N}-2)} \\
\frac{1}{3} & -\frac{2}{3} & \frac{2 (\text{N}-4)}{3 \text{N}} & \frac{\text{N}^2-6 \text{N}+11}{3 \text{N}^2-9 \text{N}+6} & -\frac{\text{N}-4}{3 (\text{N}-2)} & \frac{1}{3} \\
\frac{1}{2} & 0 & -\frac{4}{\text{N}} & \frac{-\text{N}^2+3 \text{N}+4}{2 \text{N}^2-6 \text{N}+4} & \frac{1}{2} & -\frac{1}{2} \\
\frac{1}{6} & \frac{2}{3} & -\frac{2 (\text{N}+2)}{3 \text{N}} & \frac{\text{N}^2+3 \text{N}+2}{6 \left(\text{N}^2-3 \text{N}+2\right)} & \frac{\text{N}+2}{12-6 \text{N}} & \frac{1}{6} \\
\end{array}
\right)
\end{eqnarray}
The matrices $\sigma^\mu_{p,q}$ used in \eqref{crossingPsona} is defined as
\begin{eqnarray}\label{Ysonadef}
\sigma^S_{p,q}&=& \left(\sum^L_{\ell=0, \text{even}}4\frac{\gamma'_S(-\ell+p-1,\ell) J^{\Delta -1}(-\ell+p-1,q)-\gamma'_S(p,\ell) J^{\Delta -1}(\ell+p+1,q)}{N(N-1)} \right)\nonumber\\
\sigma^F_{p,q}&=& \left(\sum^L_{\ell=1, \text{odd}}-\frac{\gamma'_F(-\ell+p-1,l) J^{\Delta -1}(-\ell+p-1,q)-\gamma'_F(p,\ell) J^{\Delta -1}(\ell+p+1,q)}{N-2})\ \right)\nonumber\\
\sigma^T_{p,q}&=& \left(\sum^L_{\ell=0, \text{even}}\frac{\gamma'_T(-\ell+p-1,l) J^{\Delta -1}(-\ell+p-1,q)-\gamma'_T(p,\ell) J^{\Delta -1}(\ell+p+1,q)}{N-2} \right)\nonumber\\
\sigma^R_{p,q}&=& 2\left(\sum^L_{\ell=0, \text{even}}\frac{\gamma'_R(-\ell+p-1,l) J^{\Delta -1}(-\ell+p-1,q)-\gamma'_R(p,\ell) J^{\Delta -1}(\ell+p+1,q)}{3} \right)\nonumber\\
\sigma^{Ms}_{p,q}&=& -\left(\sum^L_{\ell=1, \text{odd}}\gamma'_{Ms}(-\ell+p-1,l) J^{\Delta -1}(-\ell+p-1,q)-\gamma'_{Ms}(p,\ell) J^{\Delta -1}(\ell+p+1,q) \right)\nonumber\\
\sigma^{A}_{p,q}&=& \left(\sum^L_{\ell=0, \text{even}}\frac{\gamma'_{A}(-\ell+p-1,l) J^{\Delta -1}(-\ell+p-1,q)-\gamma'_{A}(p,\ell) J^{\Delta -1}(\ell+p+1,q)}{3} \right)\nonumber\\
\end{eqnarray}

where
\begin{eqnarray}
\gamma'_i&=& \frac{2(\ell+1)(2\D+2n+\ell-2)}{(\D-1)^2}\gamma^{(i)}(n,\ell)\nonumber\\
\end{eqnarray}
and $J^{\Delta -1}(p,q)$ is defined as,

\begin{eqnarray}\label{jdefn}
J^{\Delta-1}(m,m')&=& \frac{C_m}{C_{m'}} I(m,m'),\qquad I(m,m')=\oint_C \frac{dz}{2\pi i} \frac{(1-z)^m}{z^{m'+1}} F_{\D-1 + m} (z) F_{2-\D-m'}(z)\nonumber\\
C_p &=& \frac{\Gamma (p+2 (\Delta -1)-1) \Gamma (p+\Delta -1)^2}{p! \Gamma (\Delta -1)^2 \Gamma (2 p+2 (\Delta -1)-1)}\nonumber\\
\end{eqnarray}
In \cite{Heemskerk:2009pn, Heemskerk:2010ty}, the closed form expressions for $J^{\D-1}(m,m')$ were also given 
\begin{eqnarray}\label{jexplicit}
J^{\Delta-1}(m,m')&=& -\frac{\Gamma (2 p+2 \Delta-2 ) C_p}{\Gamma (p+\Delta-1 )^2 C_q}\sum _{\ell =0}^p \frac{(-1)^{\ell } \binom{p}{\ell } \left((p+\Delta-1 )_{q-\ell }\right){}^2}{\Gamma (q-\ell +1)^2}\nonumber\\
&& \, _4F_3(\ell -q,\ell -q,-q-\Delta +2,-q-\Delta +2;-p-q+\ell -\Delta +2,\nonumber\\
&&\phantom{\, _4F_3(\ell -q,\ell -q,-q-\Delta +2,-q-\Delta +2;}-p-q+\ell -\Delta +2,-2 q-2 \Delta +4;1)\nonumber\\
\end{eqnarray}


The crossing matrix corresponding to \eqref{crossingPsunf} are given by, 
\begin{eqnarray}\label{crossingmat1sunf}
M_{SU(N),f}=\left(
\begin{array}{cc}
\frac{1}{N} & 1-\frac{1}{N^2}\\ 
1 & -\frac{1}{N}\\
\end{array}
\right)
\end{eqnarray}
\begin{eqnarray}\label{crossingmat2sunf}
\tilde{M}_{SU(N),f}=\left(
\begin{array}{cc}
1+\frac{1}{N} & 1-\frac{1}{N}\\ 
1 & -1\\
\end{array}
\right)
\end{eqnarray}

The matrices  $\kappa^\mu_{p,q}$ and $\tau^\mu_{p,q}$ used in \eqref{crossingPsunf}
are given by, 
\begin{eqnarray}\label{Ysunfdef}
\kappa^{(S)}_{p,q}&=&\frac{\sum _{\ell=0}^L \left(\gamma _s(-l+p-1,l) J(\Delta -1,-l+p-1,q)-\gamma _s(p,l) J(\Delta -1,l+p+1,q)\right)}{N}\nonumber\\
\kappa^{(Adj)}_{p,q}&=&\frac{\sum _{\ell=0}^L \left(\gamma _s(-l+p-1,l) J(\Delta -1,-l+p-1,q)-\gamma _s(p,l) J(\Delta -1,l+p+1,q)\right)}{N}\nonumber\\
\tau^{(S)}_{p,q}&=&\frac{\sum _{\ell=0}^L (-1)^\ell\left(\gamma _s(-l+p-1,l) J(\Delta -1,-l+p-1,q)-\gamma _s(p,l) J(\Delta -1,l+p+1,q)\right)}{N}\nonumber\\
\tau^{(Adj)}_{p,q}&=&\frac{\sum _{\ell=0}^L (-1)^\ell\left(\gamma _s(-l+p-1,l) J(\Delta -1,-l+p-1,q)-\gamma _s(p,l) J(\Delta -1,l+p+1,q)\right)}{N}\nonumber\\
\Omega^{(Sym)}_{p,q}&=&\frac{\sum _{\ell=0, \text{even}}^L \left(\gamma _s(-l+p-1,l) J(\Delta -1,-l+p-1,q)-\gamma _s(p,l) J(\Delta -1,l+p+1,q)\right)}{N}\nonumber\\
\Omega^{(Anti-sym)}_{p,q}&=&\frac{\sum _{\ell=1, \text{odd}}^L \left(\gamma _s(-l+p-1,l) J(\Delta -1,-l+p-1,q)-\gamma _s(p,l) J(\Delta -1,l+p+1,q)\right)}{N}\nonumber\\
\end{eqnarray}

The crossing matrix $M_{SU(N),a}$ defined in \eqref{crossingsuna} is given by,
\begin{eqnarray}\label{crossingmatsuna} 
M_{SU(N),a}&=&\left(
\begin{array}{cccccc}
\frac{1}{\text{N}^2-1} & \frac{2 \text{N}}{\text{N}^2-1} & \frac{8-2 \text{N}^2}{\text{N}-\text{N}^3} & \frac{\text{N}^2-4}{\text{N}^2-1} & \frac{(\text{N}-3) \text{N}^2}{(\text{N}-1)^2 (\text{N}+1)} & \frac{\text{N}^2 (\text{N}+3)}{(\text{N}-1) (\text{N}+1)^2} \\
\frac{1}{2 \text{N}} & \frac{1}{2} & \frac{1}{2}-\frac{2}{\text{N}^2} & 0 & \frac{1}{1-\text{N}}+\frac{1}{2} & -\frac{\text{N}+3}{2 \text{N}+2} \\
\frac{\text{N}}{2 \left(\text{N}^2-4\right)} & \frac{\text{N}^2}{2 \left(\text{N}^2-4\right)} & \frac{\text{N}^2-12}{2 \left(\text{N}^2-4\right)} & \frac{\text{N}}{4-\text{N}^2} & -\frac{(\text{N}-3) \text{N}^3}{2 (\text{N}-2)^2 \left(\text{N}^2+\text{N}-2\right)} & \frac{\text{N}^3 (\text{N}+3)}{2 (\text{N}-2) (\text{N}+1) (\text{N}+2)^2} \\
\frac{1}{2} & 0 & -\frac{2}{\text{N}} & \frac{1}{2} & \frac{1}{\text{N}-2}+\frac{1}{1-\text{N}}-\frac{1}{2} & -\frac{1}{\text{N}+2}+\frac{1}{\text{N}+1}-\frac{1}{2} \\
\frac{1}{4} & \frac{1}{2} & -\frac{\text{N}+2}{2 \text{N}} & -\frac{\text{N}+2}{4 \text{N}} & \frac{1}{2-2 \text{N}}+\frac{1}{\text{N}-2}+\frac{1}{4} & \frac{\text{N}+3}{4 \text{N}+4} \\
\frac{1}{4} & -\frac{1}{2} & \frac{1}{2}-\frac{1}{\text{N}} & -\frac{\text{N}-2}{4 \text{N}} & \frac{\text{N}-3}{4 (\text{N}-1)} & \frac{\text{N}^2+\text{N}+2}{4 \text{N}^2+12 \text{N}+8} \\
\end{array}
\right)
\end{eqnarray}
The matrices $\Lambda^\mu_{p,q}$ used in \eqref{crossingPsuna} is defined as
\begin{eqnarray}\label{Ysunadef}
\Lambda^S_{p,q}&=&\sum_{\ell=0, \text{even}}^L 2\left(\frac{\gamma'_S(-l+p-1,l) J(\Delta -1,-l+p-1,q)-\gamma'_S(p,l) J(\Delta -1,l+p+1,q)}{(N+1)(N-1)}\right)\nonumber\\
\Lambda^{Adj_{-}}_{p,q}&=&\sum_{\ell=1, \text{odd}}^L -\left(\frac{\gamma'_{Adj_{-}}(-l+p-1,l) J(\Delta -1,-l+p-1,q)-\gamma'_{Adj_{-}}(p,l) J(\Delta -1,l+p+1,q)}{N}\right)\nonumber\\
\Lambda^{Adj_{+}}_{p,q}&=&\sum_{\ell=0, \text{even}}^L N\left(\frac{\gamma'_{Adj_{-}}(-l+p-1,l) J(\Delta -1,-l+p-1,q)-\gamma'_{Adj_{-}}(p,l) J(\Delta -1,l+p+1,q)}{(N+2)(N-2)}\right)\nonumber\\
\Lambda^{AS}_{p,q}&=&\sum_{\ell=1, \text{odd}}^L -\left(\gamma'_{AS}(-l+p-1,l) J(\Delta -1,-l+p-1,q)-\gamma'_{AS}(p,l) J(\Delta -1,l+p+1,q)\right)\nonumber\\
\Lambda^{AA}_{p,q}&=&\sum_{\ell=0, \text{even}}^L \left(\frac{\gamma'_{AA}(-l+p-1,l) J(\Delta -1,-l+p-1,q)-\gamma'_{AA}(p,l) J(\Delta -1,l+p+1,q)}{2}\right)\nonumber\\
\Lambda^{SS}_{p,q}&=&\sum_{\ell=0, \text{even}}^L \left(\frac{\gamma'_{SS}(-l+p-1,l) J(\Delta -1,-l+p-1,q)-\gamma'_{SS}(p,l) J(\Delta -1,l+p+1,q)}{2}\right)\nonumber\\
\end{eqnarray}

where
\begin{eqnarray}
\gamma'_i&=& \frac{2(\ell+1)(2\D+2n+\ell-2)}{(\D-1)^2}\gamma^{(i)}(n,\ell)\nonumber\\
\end{eqnarray}
and $J^{\Delta -1}(p,q)$ has been defined in \eqref{jdefn}. The matrices ${\b'}^\mu_{p,q}$ and $M_{SO(4),f}$ used in \eqref{crossingPso4f} are defined as 

\begin{eqnarray}\label{cmson4}
M_{SO(4),f}=\left(
\begin{array}{cccc}
\frac{1}{4} & \frac{9}{16} & -\frac{3}{16} & -\frac{3}{16} \\
1 & \frac{1}{4} & \frac{1}{4} & \frac{1}{4} \\
-1 & \frac{3}{4} & -\frac{1}{4} & \frac{3}{4} \\
-1 & \frac{3}{4} & \frac{3}{4} & -\frac{1}{4} \\
\end{array}
\right)
\end{eqnarray}
\begin{eqnarray}\label{pso4fdef}
{\b'}^S_{p,q}&=& \left(\sum^L_{\ell=0, \text{even}}\frac{\gamma'_S(-\ell+p-1,\ell) J^{\Delta -1}(-\ell+p-1,q)-\gamma'_S(p,\ell) J^{\Delta -1}(\ell+p+1,q)}{n}) \right)\nonumber\\
{\b'}^T_{p,q}&=& \left(\sum^L_{\ell=0, \text{even}}\frac{\gamma'_T(-\ell+p-1,l) J^{\Delta -1}(-\ell+p-1,q)-\gamma'_T(p,\ell) J^{\Delta -1}(\ell+p+1,q)}{n}) \right)\nonumber\\
{\b'}^A_{p,q}&=& -\left(\sum^L_{\ell=1, \text{odd}}\frac{\gamma'_A(-\ell+p-1,l) J^{\Delta -1}(-\ell+p-1,q)-\gamma'_A(p,\ell) J^{\Delta -1}(\ell+p+1,q)}{n}) \right)\nonumber\\
{\b'}^{A'}_{p,q}&=& -\left(\sum^L_{\ell=1, \text{odd}}\frac{\gamma'_{A'}(-\ell+p-1,l) J^{\Delta -1}(-\ell+p-1,q)-\gamma'_{A'}(p,\ell) J^{\Delta -1}(\ell+p+1,q)}{n}) \right)\nonumber\\
\end{eqnarray}

\section{Relation between $\hat{\beta}_{i,j}$  and $\beta_{n,\ell}$ } \label{relnbetwnbb}

Here we briefly discuss the relation between  $\hat{\beta}_{i,j}$ and $\beta_{i,j}$ following \cite{Mazac:2019shk}. Let's define the following generating functional,
\begin{equation}
\sum_{n,\ell} G_{\Delta_B,\ell}(z,\bar{z}) \alpha_{n,\ell}+\sum_{n,\ell} \partial_{\Delta}G_{\Delta_B,\ell}(z,\bar{z}) \beta_{n,\ell}
\end{equation}
where $\Delta_B=2\Delta_{\phi}+2n+\ell$. Using the explicit form of conformal block \cite{Dolan:2003hv,Dolan:2011dv, Hogervorst:2013sma},
\begin{equation}
G_{\Delta,\ell}=\sum_{n=0}^{\infty}\sum_j A_{n,j} \mathcal{P}_{\Delta+n,J},
\end{equation}
where,
\begin{equation}
\mathcal{P}_{E,J}(s,\xi)=s^E \frac{j!}{(2\nu)_j} C_j^{\nu}(\xi).
\end{equation}

The coefficients $A_{n,j}$ satisfy the following recursion relation,
\begin{equation}
\left(C_{\Delta+n,j}-C_{\Delta,\ell}\right)A_{n,j} =\gamma^+_{\Delta+n-1,j-1}A_{n-1,j-1}+\gamma^-_{\Delta+n-1,j+1}A_{n-1,j+1}
\end{equation}
where,
\begin{equation}
\gamma^+_{E,J}=\frac{(E+j)^2 (j+2\nu)}{2(j+\nu)},\,\,\, \gamma^-_{E,J}=\frac{(E-j-2\nu)^2 j}{2(j+\nu)}
\end{equation}
and we have the initial conditions $A_{0,j}=\delta_{jl}$. Using this the generating functional will take the following form,
\begin{equation}
\sum_{i,j=0}^{\infty} \left(\hat{\alpha}_{i,j}+\hat{\beta}_{i,j}\frac{\log(z\bar{z})}{2}\right) z^i \bar{z}^j
\end{equation}

$\hat{\alpha}_{i,j}$,$\hat{\beta}_{i,j}$ are independent of spacetime dimension and dimension of external operator. These are easy to compute and using the expansion of conformal block we can always translate $\hat{\alpha}_{i,j}$,$\hat{\beta}_{i,j}$ to $\alpha_{n,\ell}$,$\beta_{n,\ell}$.

\section{Scattering of four majorana fermions in 1+1 dimensions}\label{majscat}
\paragraph{ } In this appendix we set up the scattering kinematics and the necessary ingredients for massive majorana fermion scattering in 1+1 dimensions. In particular, we review the 1+1 dimensional scattering kinematics and also provide the explicit expressions for S-matrix for the Lagrangians listed in subsection \ref{mffssi11d}. Our results are consistent with the analysis of \cite{Jain:2014nza} when restricted to 1+1 dimensions. 

\subsection*{Kinematics}
Consider the scattering of four identical massive particles in $2$-dimensional Minkowski space. Let $p_i^\mu$ be momentum of the $i^{\rm th}$ particle with mass $m$. Momentum conservation and on-shell condition implies
\begin{equation}\label{osmc}
p_i^2=-m^2,\qquad p_1^\mu+p_2^\mu -p_3^\mu-p_4^\mu=0.
\end{equation}
For convenience we parametrize the momenta in the following manner,
\begin{equation}
p_i^\mu=(m\cosh \alpha_i,~ \sinh \alpha_i).
\end{equation}
We use the convention that particles with momenta $p_1$ and $p_2$ are incoming and $p_3$ and $p_4$ are outgoing. Momentum conservation implies $p_1=p_4,~ p_2=p_3$. The Mandelstam variables can be defined as follows,
\begin{equation}\label{stu} \begin{split}
s&:=-(p_1+p_2)^2=-(p_3+p_4)^2=2m^2-2 p_1.p_2,\\
t&:=-(p_1-p_3)^2=-(p_4-p_2)^2= 2m^2+2 p_1.p_3,\\
u&:=-(p_1-p_4)^2=-(p_2-p_3)^2=0.
\end{split}
\end{equation}
The equalities in \eqref{stu} follow from \eqref{osmc}. Due to momentum conservation the Mandelstam invariants are related by  $s+t=4m^2$. Contrary to $D\geq3$, for the special case $D=2$, the kinematics of four particle scattering degenerates and variables  $u$ and $t$ can be solved for in terms of $s$ and $m^2$ (in particlar $u=0$). 

\subsection*{Plane wave solutions to majorana equation}
\paragraph{ } We consider the plane wave solutions to \eqref{majeqn} for majorana fermions with no colour as 

\begin{equation}
\begin{split}
\psi(x) &= \int \frac{dp}{(2\pi) 2p_0} \left( b(p) u(p) e^{ip_\mu x^\mu} + c^\dagger(p) v(p) e^{ip_\mu x^\mu} \right)
\end{split}
\end{equation}
where, the commutation relations and the majorana condition \eqref{majoranacondition} implies 
\begin{equation}
c(p) = b(p), \qquad u^\star(p) =v(p),\qquad \{b(p),b^\dagger(p')\}= 2\pi \delta(p-p')2p_0
\end{equation}
The momentum space majorana equation becomes,
\begin{eqnarray}\label{momspmaj}
(ip_\mu\g^\mu-m)u(p)&=&0,\qquad (-ip_\mu\g^\mu-m)v(p)=0\nonumber\\
\bar{u}(p)(ip_\mu\g^\mu-m)&=&0,\qquad \bar{v}(p)(-ip_\mu\g^\mu-m)=0
\end{eqnarray}
The general procedure for evaluating this involves going to the rest frame and then boosting the solution obtained. This procedure gives the following solutions to $u(p)$ and $v(p)$,

\begin{eqnarray}\label{uvmaj}
u(p_i)=\sqrt{m}(e^{-\a_i/2},i e^{\a_i/2}),\qquad v(p_i)=\sqrt{m}(e^{-\a_i/2},-i e^{\a_i/2})
\end{eqnarray}
We then evaluate the relevant inner products as,
\begin{eqnarray} 
\bar{u}(p_1)\cdot u(p_2)&=& \sqrt{2m^2-2p_1\cdot p_2},\qquad \bar{v}(p_1)\cdot v(p_2)= -\sqrt{2m^2-2p_1\cdot p_2} \nonumber\\
\bar{u}(p_1)\cdot v(p_2)&=& \sqrt{-2m^2-2p_1\cdot p_2},\qquad \bar{v}(p_1)\cdot u(p_2)= -\sqrt{-2m^2-2p_1\cdot p_2} \nonumber\\
\bar{u}(p_2)\cdot v(p_1)&=& -\sqrt{-2m^2-2p_1\cdot p_2},\qquad \bar{v}(p_2)\cdot u(p_1)= \sqrt{-2m^2-2p_1\cdot p_2} \nonumber\\
\end{eqnarray} 
where $\bar{A}=A^T C$ as defined in \eqref{majoranacondition}. 

\subsection*{Fermion S-matrix}
                 In this subsection we write the explicit form of the fermion S-matrix 
which is generated by the local Lagrangians listed in \eqref{fermlagnocol} and \eqref{fermlagsonfund} respectively. Consider the scattering process of four identical majorana fermions
$$\psi_\a + \psi_\b \rightarrow \psi_\g + \psi_\d$$
 where the subscript denotes the spinor index. 
 Using the plane wave solutions listed in \eqref{uvmaj}, the fermion S-matrix can be written as

\begin{eqnarray}\label{smatfermnocol}
S^{L^{F}}&=&(\bar{u}(p_3)\cdot u(p_2) \bar{u}(p_4)\cdot u(p_1)+\bar{v}(p_2)\cdot v(p_3) \bar{v}(p_1)\cdot v(p_4)) (p_3\cdot p_2)^m(-p_3\cdot p_4)^n\nonumber\\
&& + (\bar{u}(p_3)\cdot u(p_1) \bar{v}(p_2)\cdot v(p_4)+\bar{v}(p_1)\cdot v(p_3) \bar{u}(p_4)\cdot u(p_2))(p_3\cdot p_2)^m(p_3\cdot p_1)^n\nonumber\\
&& + (\bar{u}(p_3)\cdot v(p_4) \bar{v}(p_1)\cdot u(p_2)+ \bar{v}(p_4)\cdot u(p_3)\bar{u}(p_2)\cdot v(p_1))(p_3\cdot p_1)^m(-p_3\cdot p_4)^n\nonumber\\
\end{eqnarray} 

Consider the scattering process of four identical majorana fermions charged under the fundamental of $SO(N)$,
$$\psi^a_\a + \psi^b_\b \rightarrow \psi^c_\g + \psi^d_\d$$
where the superscript denotes the internal symmetry index while the subscript denotes the spinor index as usual.
Using the plane wave solutions listed in \eqref{uvmaj}, the fermion S-matrix can be written as

\begin{eqnarray}\label{smatfermsonfund}
S^{L_{SO(N),f}^{F}}&=&(\bar{u}(p_3)\cdot u(p_2) \bar{u}(p_4)\cdot u(p_1)+\bar{v}(p_2)\cdot v(p_3) \bar{v}(p_1)\cdot v(p_4))(p_3\cdot p_2)^m(-p_3\cdot p_4)^n \delta_{bc}\delta_{ad}\nonumber\\
&& + (\bar{u}(p_3)\cdot u(p_1) \bar{v}(p_2)\cdot v(p_4)+\bar{v}(p_1)\cdot v(p_3) \bar{u}(p_4)\cdot u(p_2))(p_3\cdot p_2)^m(p_3\cdot p_1)^n \delta_{ac}\delta_{bd}\nonumber\\
&& + (\bar{u}(p_3)\cdot v(p_4) \bar{v}(p_1)\cdot u(p_2)+ \bar{v}(p_4)\cdot u(p_3)\bar{u}(p_2)\cdot v(p_1))(p_3\cdot p_1)^m(-p_3\cdot p_4)^n \delta_{ab}\delta_{cd}\nonumber\\
\end{eqnarray}

\section{Plethystic Integrals}\label{PI}
In this appendix we detail the evaluation of colour integral of \eqref{singlet-proj} and \eqref{singlet-proj-ferm}

\bea
I^R_\ts(x):=\oint  d\mu_{G}~ \oint  d\mu_{SO(D)}~  i_\ts^{(4)}(x,y,z)/\denom(x,y),\qquad 
I^R_\tf(x):=\oint  d\mu_{G}~ \oint  d\mu_{SO(D)}~  i_\tf^{(4)}(x,y,z)/\denom(x,y).\nonumber\\
\eea

where $d\mu_{SO(D)}$ is the haar measure associated with the Lorentz group $SO(D)$ and $d\mu_{G}$ is the haar measure associated with the colour group $G$. The haar measure for $SO(D)$ for even dimensions ($D=2N$) and odd dimensions ($D=2N+1$)are  given by, 

\begin{eqnarray}\label{haare}
\Delta_e(y_i) =\frac{2 \left(\prod _{j=1}^{N} \left(\prod _{i=1}^{j-1} \left(y_i+\frac{1}{y_i}-y_j-\frac{1}{y_j}\right)\right)\right)^2}{(2\pi i)^N 2^NN!\prod_{i=1}^{N}y_i},\nonumber\\
\Delta_o(y_i) =\frac{\left(\prod _{k=1}^N \left(1-y_k-\frac{1}{y_k}\right)\right) \left(\prod _{j=1}^N \left(\prod _{i=1}^{j-1}  \left(y_i+\frac{1}{y_i}-y_j-\frac{1}{y_j}\right)\right)\right)^2}{(2\pi i)^N N!\prod_{i=1}^{N}y_i}
\end{eqnarray}
and the integral over $y_i$ in \eqref{singlet-proj} is a closed circular contour about $y_i=0$. 
The Haar measure for $SU(N)$ is given by \cite{Gray:2008yu}, 
\be\label{haarsu}
d\mu_{SU(N)}=\frac{1}{(2\pi i)^{(N-1)} N!} \prod_{l=1}^{N-1} \frac{dz_l}{z_l} \D(\phi)\D(\phi^{-1})	
\ee 
where $\phi_a(z_1,\ldots z_{N-1})|_{a=1}^N$ are the coordinates on the maximal torus of $SU(N)$ with $\prod_{l=1}^N \phi_l=1$ and $\D(\phi)=\prod_{1\leq a< b\leq N} (\phi_a-\phi_b)$ is the Vandermonde determinant. Similar to $SO(D)$, the integral over $z_i$ in \eqref{singlet-proj} is a closed circular contour about $z_i=0$. Explicitly written out, the coordinates on the maximal torus take the form, 
\be\label{coordmaxtor}
\phi_1=z_1,\qquad \phi_k=z^{-1}_{k-1}z_k,\qquad \phi_N=z^{-1}_{N-1}
\ee

We, therefore, have to perform two Haar integrals one of which projects onto Lorentz singlets, while the other projects onto the colour singlets. We perform the Haar integral for the Lorentz singlets first using the Large $D$ techniques used in \cite{Chowdhury:2019kaq}, keeping in mind that the Haar integral for the lorentz group stabilizes for $D>3$ for scalars. We obtain the following general result

\begin{eqnarray}
	I^D_\ts(x):= \oint d\mu_{G}~&\left(\frac{\chi^{G}_R(z^2) \chi^{G}_R(z)^2}{4 \left(1-x^4\right)}+\frac{\chi^{G}_R(z^4)}{4 \left(1-x^4\right)}+\frac{\chi^{G}_R(z)^4}{24 \left(x^2-1\right)^2}+\frac{\chi^{G}_R(z^2)^2}{8 \left(x^2-1\right)^2}\right.\nonumber\\
	&\left. +\frac{\chi^{G}_R(z^3)\chi^{G}_R(z)}{3 \left(x^4+x^2+1\right)}\right)
\end{eqnarray}

Now let us consider the colour projections case by case. For the cases where the scalar field is transforming in the fundamental and adjoint of $SO(N)$ and $SU(N)$, we expect the final partition function to be a sum of the partition functions $Z$ of representations of $S_3$  (see subsection 2.9 of \cite{Chowdhury:2019kaq}). More precisely we expect, 

\begin{equation} \label{pfnnnn}
Z_{ \text{S-matrix}}(x) = \sum_{J} x^{\Delta_J} 
Z_{{\bf R_J}}(x)
\end{equation} 
where $Z_{\bf R_J}(x)$ are listed in \eqref{partfnsss}.

\begin{equation} \label{partfnsss}
\begin{split} 
&Z_{{\bf 1_S}}(x)=\denom, \qquad Z_{{\bf 1_A}}(x)=x^{6}\denom
,\qquad Z_{{\bf 2_M}}(x)=(x^2+x^4)\denom, \\
& Z_{{\bf 3}}(x)=
Z_{{\bf 1_S}} + Z_{{\bf 2_M}}(x)=\left( 1+ x^2 +x^4 \right) \denom \\
& Z_{{\bf 3_A}}(x)=
Z_{{\bf 1_A}} + Z_{{\bf 2_M}}(x)=\left( x^6+ x^2 +x^4 \right) \denom \\
&Z_{{\bf 6}}(x)=
Z_{{\bf 1_S}}+ Z_{{\bf 1_A}}  + 2Z_{{\bf 2_M}}(x)=\left( 1+ 2x^2 +2x^4 +x^6 \right) \denom \\
&{\rm where}\quad \denom=\frac{1}{(1-x^4)(1-x^6)} \\
\end{split}
\end{equation} 

\subsection{Integrals for $SO(N)$ fundamental and adjoint}
In this subsection, we set up the computation for the cases when $G=SO(N)$ and $R$ is the fundamental and the adjoint representation of $SO(N)$. Since we are interested in large $N$ computation, it suffices to perform the Haar integral over colour singlets using the large $N$ techniques of \cite{Chowdhury:2019kaq, Chowdhury:2020ddc}. We obtain the following results for fundamental and adjoint representation of $SO(N)$ (see table \ref{sofund}).

\begin{table}[h!]
	\begin{center}
		\begin{tabular}{|l|c|r|} 
			\hline
			\textbf{Integral} & \textbf{$SO(N)$ fundamental} & \textbf{$SO(N)$ adjoint}\\
			\hline
			$\oint \prod_{i=1}^{\lfloor N/2\rfloor}dy_i \,\Delta(y_i)\,~ \chi^{SO(N)}_R( y^2)\chi^{SO(N)}_R( y)^2$& 1 & 2\\
			\hline 
			$\oint \prod_{i=1}^{\lfloor N/2\rfloor}dy_i \,\Delta(y_i)\, ~\chi^{SO(N)}_R( y^4)$ & 1 & 2\\
			\hline 
			$\oint \prod_{i=1}^{\lfloor N/2\rfloor}dy_i \,\Delta(y_i)\, ~\chi^{SO(N)}_R( y)^4$ & 3 & 6\\
			\hline 
			$\oint \prod_{i=1}^{\lfloor N/2\rfloor}dy_i \,\Delta(y_i)\, ~\chi^{SO(N)}_R( y^2)^2$ & 3 & 6\\
			\hline 
			$
			\oint \prod_{i=1}^{\lfloor N/2\rfloor}dy_i \,\Delta(y_i)\, ~\chi^{SO(N)}_R( y^3)\chi^{SO(N)}_R( y) $ & 0 & 0\\
			\hline
		\end{tabular}
		\caption{Large $N$ integrals for SO(N)}\label{sofund}
	\end{center}
\end{table}

	
	where  
	\bea 
	\chi^{SO(N)}_f( y) &=& \sum_{i=1}^{\lfloor N/2\rfloor}\left( y_i + \frac{1}{y_i} \right)\nonumber\\
	\chi^{SO(N)}_a( y) &=& \left( \frac{\chi^{SO(N)}_f(y)^2-\chi^{SO(N)}_f(y^2)}{2} \right).
	\eea

For low $N$, the integrals can be done using numerical integrations outlined in \cite{Chowdhury:2019kaq, Chowdhury:2020ddc}. We list the results of the haar integral in table \ref{sofund}. 
\subsection{Integrals for $SU(N)$ adjoint}	
	
The results for the adjoint representation has already been listed in \cite{Chowdhury:2020ddc} and reproduced here for convenience.


 	\begin{table}[h!]
 		\begin{center}
 		\begin{tabular}{|l|c|} 
 				\hline
 				\textbf{Integral} & \textbf{$SU(N)$ adjoint}\\
 				\hline
 				$	\frac{1}{(2\pi i)^{(N-1)} N!}\oint  \prod_{l=1}^{N-1} \frac{dz_l}{z_l} ~ \chi^{SU(N)}_a( z^2)\chi^{SU(N)}_a( z)^2$& 1 \\
 				\hline 
 				$	\frac{1}{(2\pi i)^{(N-1)} N!}\oint  \prod_{l=1}^{N-1} \frac{dz_l}{z_l} ~ \chi^{SU(N)}_a( z^4)$ & 3 \\
 				\hline 
 				$\frac{1}{(2\pi i)^{(N-1)} N!}\oint  \prod_{l=1}^{N-1} \frac{dz_l}{z_l} ~ \chi^{SU(N)}_a( z)^4$& 9 \\
 				\hline
 				$\frac{1}{(2\pi i)^{(N-1)} N!}\oint  \prod_{l=1}^{N-1} \frac{dz_l}{z_l} ~ \chi^{SU(N)}_a( z^2)^2$ & 5\\
 				\hline 
 				$\frac{1}{(2\pi i)^{(N-1)} N!}\oint  \prod_{l=1}^{N-1} \frac{dz_l}{z_l} ~ \chi^{SU(N)}_a( z)\chi^{SU(N)}_a( z^3)$ & 0\\
 				\hline
 			\end{tabular}
 			\caption{Large $N$ integrals for SU(N) adjoint}\label{suadj}
 		\end{center}
 	\end{table}

	where 
	
	\bea
	\chi^{SU(N)}_f\left(z\right)&=& \sum_{i=1}^N z_i, \qquad
	\chi^{SU(N)}_{\bar{f}}\left( z\right)= \sum_{i=1}^N z_i^{-1}  \nonumber\\
	\chi^{SU(N)}_a \left(z \right) &=& \chi^{SU(N)}_f\left(z \right) \chi^{SU(N)}_{\bar{f}} \left(z \right) -1
	\eea   
	

\subsection{Fermions}    
In this subsection we present the Haar integral over the $D=2$ space-time for fermions. Quoting \eqref{singlet-proj-ferm},
\be
\begin{split}
	I^R_\tf(x)&=\oint  d\mu_{G}~ \oint  d\mu_{SO(2)}~ \frac{1}{24}\Big(i^4_\tf(x,y,z) -6 i^2_\tf(x,y,z) i_\tf(x^2,y^2,z^2)+3i^2_\tf(x^2,y^2,z^2)\\&+8i_\tf(x,y,z)i_\tf(x^3,y^3,z^3)-6i_\tf(x^4,y^4,z^4)\Big).
\end{split}
\ee  
The lorentz integral was done numerically while the colour integral has been done using the large $N$ techniques explained in the previous section. An important difference from the previous section is the following fact that, in $D=2$, the number of invariant mandelstam invariants is just one instead of two for higher dimensions. This implies the final result can be written in the form 
\begin{equation} 
Z^{D=2}_{ \text{S-matrix}}(x) = \sum_{J} x^{\Delta_J} 
Z_{{\bf R_J}}(x)
\end{equation}  
where $Z_{{\bf R_J}}(x)$ are the partition functions corresponding to the irreducible representations of $S_2$ as outlined in appendix \ref{rtosami}.
\begin{eqnarray}  
{\bf Z_S}(x)=\frac{1}{1-x^4},\qquad {\bf Z_A}(x)=\frac{x^2}{1-x^4}
\end{eqnarray}

We obtained \eqref{fermionpleth} and \eqref{fermionplethcolour} in this manner.

\section{Representation theory of $S_2$ and action on mandelstam invariants}\label{rtosami}
We present the representation theory for two dimensional discrete group $S_2$ and its action on the mandelstam invariants. The permutation group of two elements has two one-dimensional irreducible representations- the totally symmetric representation and the totally anti-symmetric representation. The generator for the $S_2$ is a $\Z_2$ flip and it is almost immediately obvious the two irrducible represenations can be labelled by their $\Z_2$ charges. The standard young's diagrams associated with these representations are 

\begin{equation} \label{yts}
{\bf Z_S}={\tyng(2)}\qquad\qquad {\bf Z_A}={\tyng(1,1)}.
\end{equation}
To be precise, denoting the $\Z_2$ generator by $P_{12}$, the representation ${\bf Z_S}$ has the charge ``$+$" while $ {\bf Z_A}$ has the charge`` $-$" under the actions of $P_{12}$. We now construct the action of $S_2$ on the mandelstam invariants $s,t$ subject to the constraint $s+t=0$ (we focus on the massless case first, the massive case is a trivial generalisation of this case).  Before trying to construct polynomials that transform in the two irreducible representations of $S_2$, we first present the partition function counting. Following \cite{Chowdhury:2019kaq}, we can define the single variable partition function as 
\be
z(x):={\rm Tr}\, x^{2\Delta}=\frac{1}{1-x^2}.
\ee
Here $\Delta$ is the degree of momentum homogeneity. The partition functions over polynomials of two variables with given transformation property and the constraint $s+t=0$ are
\begin{eqnarray} \label{defz} 
{\bf Z_S}(x)&:=&\frac{1}{2}\left( z(x)^2+ z(x^2)\right)(1-x^2)=\frac{1}{1-x^4}\nonumber\\
{\bf Z_A}(x)&:=&\frac{1}{2}\left( z(x)^2- z(x^2)\right)(1-x^2)=\frac{x^2}{1-x^4}
\end{eqnarray}
The modules built out of polynomials of mandelstam invariants, corresponding to these partition functions can also be constructed as follows
\begin{eqnarray}\label{s2mandmod}
M_{\bf Z_S}(n)=s^{2n},\qquad M_{\bf Z_A}(n)=s^{2n+1}
\end{eqnarray}
where the $S_2$ properties of the modules are self-explanatory. Therefore in two dimensions, the S-matrix partition function is expected to be of the form
\begin{eqnarray}
	Z_{\text{S-matrix}}=\sum_{\Delta_J}x^{\Delta_J} Z_{\Delta_J}(x)
\end{eqnarray}
where $Z_{\Delta_J}(x)$ can be either one of \eqref{defz}. For the massive case the modules become 
\begin{eqnarray}\label{s2mandmodmassive}
M'_{\bf Z_S}(n)=(s(4m^2-s))^{n},\qquad M'_{\bf Z_A}(n)=s(s(4m^2-s))^{n}
\end{eqnarray}

                                       

\newpage 
\bibliographystyle{JHEP}
\bibliography{notes}

\end{document}